\documentclass[journal,10pt]{IEEEtran}

\usepackage{cite}
\usepackage{amsmath}
\usepackage{amssymb}
\usepackage{adjustbox}
\usepackage{algorithm}
\usepackage{graphicx}
\graphicspath{ {images/} }

\usepackage{algpseudocode}

\usepackage{enumerate}
\usepackage{soul}  
\soulregister\cite7 

\usepackage{url}
\usepackage{subfigure}
\usepackage{multirow}
\newlength\myindent
\usepackage{tikz}
\pgfdeclarelayer{myback}
\pgfdeclarelayer{myback2}
\pgfsetlayers{background,myback2,myback,main} 

\usepackage{pgfplots}
\pgfplotsset{width=10cm,compat=newest}
\pgfplotsset{compat=newest} 
\pgfplotsset{plot coordinates/math parser=false} 
\newlength\figureheight 
\newlength\figurewidth 
\usepgfplotslibrary{units}
\usepackage{pgfplotstable}

\DeclareMathOperator*{\argmax}{argmax}

\usepackage[absolute,showboxes]{textpos}

\TPMargin{5pt}

\begin{document}
	
\TPoptions{showboxes=false}
\begin{textblock}{0.84}(0.08,0.01)    
	\noindent
	\footnotesize
\begin{center}
	This article has been accepted for publication in IEEE Transactions on Vehicular Technology. This is the author's version.\\
	Citation information: DOI 10.1109/TVT.2022.3186910
\end{center}
\end{textblock}

\TPoptions{showboxes=true}
\begin{textblock}{0.84}(0.08,0.95)    
	\noindent
	\footnotesize
	\copyright 2022 IEEE. Personal use of this material is permitted. Permission from IEEE must be 	obtained for all other uses, in any current or future media, including reprinting/republishing this material for advertising or promotional purposes, creating new collective works, for resale or redistribution to servers or lists, or reuse of any copyrighted	component of this work in other works.
\end{textblock}

\title{Scheduling Out-of-Coverage Vehicular Communications Using Reinforcement Learning}

\author{Taylan \c{S}ahin,
		Ramin Khalili, 
		Mate Boban,~\IEEEmembership{Senior~Member,~IEEE,} and
		Adam Wolisz,~\IEEEmembership{Senior~Member,~IEEE}
		
		
\thanks{ 
	Copyright \copyright 2022 IEEE. Personal use of this material is permitted. However, permission to use this material for any other purposes must be obtained from the IEEE by sending a request to pubs-permissions@ieee.org.
	
	An early version of this work was presented at the IEEE 2nd Connected and Automated Vehicles Symposium (CAVS), Honolulu, Hawaii, USA, September 2019. 
	
	T. \c{S}ahin was with Huawei Technologies Duesseldorf GmbH, 80992 Munich, Germany. He is now with the Telecommunication Networks Group, Technische Universit{\"a}t Berlin, 10587 Berlin, Germany (e-mail: taylan.sahin.1@campus.tu-berlin.de)
	
	R. Khalili and M. Boban are with Huawei Technologies Duesseldorf GmbH, 80992 Munich, Germany (e-mail: ramin.khalili@huawei.com; mate.boban@huawei.com).
	
	A. Wolisz is with the Telecommunication Networks Group, Technische Universit{\"a}t Berlin, 10587 Berlin, Germany (e-mail: adam.wolisz@tu-berlin.de).
	
}
}

{}

\maketitle

\begin{abstract}

Performance of vehicle-to-vehicle (V2V) communications depends highly on the employed scheduling approach. While centralized network schedulers offer high V2V communication reliability, their operation is conventionally restricted to areas with full cellular network coverage. In contrast, in out-of-cellular-coverage areas,  comparatively inefficient distributed radio resource management is used. To exploit the benefits of the centralized approach for enhancing the reliability of V2V communications on roads lacking cellular coverage, we propose VRLS (Vehicular Reinforcement Learning Scheduler), a centralized scheduler that proactively assigns resources for out-of-coverage V2V communications \textit{before} vehicles leave the cellular network coverage. By training in simulated vehicular environments, VRLS can learn a scheduling policy that is robust and adaptable to environmental changes, thus eliminating the need for targeted (re-)training in complex real-life environments. We evaluate the performance of VRLS under varying mobility, network load, wireless channel, and resource configurations. VRLS outperforms the state-of-the-art distributed scheduling algorithm in zones without cellular network coverage by reducing the packet error rate by half in highly loaded conditions and achieving near-maximum reliability in low-load scenarios.

\end{abstract}


\IEEEpeerreviewmaketitle

\section{Introduction}
\label{Introduction}

Cooperative awareness, defined as knowledge of the location, speed, and bearing of surrounding vehicles, forms a basis for most traffic safety and efficiency applications \cite{3gppTS22185}, \cite{3gppTS22186}. Such awareness can be best achieved in real time using vehicle-to-vehicle (V2V) communication utilizing a dedicated set of radio resources. However, achieving a collision-free transmission under a highly dynamic vehicular environment is a challenging task.

Recent work in the 3rd Generation Partnership Project (3GPP) defines a solution where resources for V2V communication are efficiently coordinated by a base station (BS) of a cellular network, resulting in highly reliable transmissions \cite{3gppTR36885}, \cite{3gppTR38885}. However, V2V transmissions should not rely completely on ubiquitous availability of the cellular network coordination, as there will always exist coverage gaps, either because of physical impediments (e.g., tunnels, blockage of a link by large objects such as buildings, etc.) or limited infrastructure deployment. 

Existing solutions for scheduling V2V communications outside the network coverage are based on distributed methods, where vehicles select resources autonomously. 3GPP defines such a resource allocation method (named ``mode 4'' in Long Term Evolution (LTE) \cite{3gppTR37885}, and ``mode 2'' in the {\color{black}fifth-generation} (5G) New Radio (NR) \cite{3gppTR37985}), which is based on sensing of the radio resources by the vehicles. Another solution is defined by the IEEE standard 802.11p \cite{ieee11p} (and its successor 802.11bd \cite{ieee11bd}) based on carrier-sensing multiple access. Although distributed solutions do not require a network infrastructure, they suffer from the hidden node problem due to the limited local view, which results in reduced quality of service \cite{vukadinovic20183gpp}, \cite{bazzi2017}, \cite{gozalvez2020}. Given that the use cases require consistently high reliability and low latency of V2V communication to be maintained irrespective of coverage \cite{3gppTR38885}, the intermittent coverage poses one of the key problems in assuring proper V2V connectivity.

In this paper, we study an alternative approach to the V2V resource scheduling in the out-of-coverage (OOC) zones. We propose a centralized scheduler residing in the network, comprising a reinforcement learning (RL) \textit{agent} that \textit{pre-assigns} resources to the vehicles for their V2V transmissions in an OOC area, before the vehicles enter such an area. {\color{black} We are motivated to use an RL-based approach due to its successful applications for resource allocation problems in general \cite{mao2016resource}, and for vehicular communications in particular \cite{magazine}.} The scheduler, {\color{black}which we call} VRLS (Vehicular Reinforcement Learning Scheduler), was proposed and demonstrated to be potentially efficient in \cite{sahin2019vrls}. VRLS design addresses two practical challenges. First, training and deploying RL solutions in the real world is costly in terms of training. In particular:
\begin{itemize}
\item Training an RL agent in a real-world setting is considerably slower than training it in a simulated environment, because of the limited availability of data samples.
\item Collecting data from an actual vehicular network is expensive, or might not be even possible considering the additional signaling and processing overhead it incurs.
\item Any undesirable outcomes of an RL agent still under training might threaten the safety-critical V2V use cases.
\end{itemize}
{\color{black}Second, in real-world problems, it is likely that the conditions in a given environment, such as road traffic mobility or data traffic load in a V2V communication network, would change over time. It would be impractical to re-design, re-train, and re-evaluate a new RL solution every time the environment changes, even if {\color{black}this} change is substantial. Therefore, a single RL-based solution should be applicable to varying conditions in the environment.}

{\color{black}By defining a unified state representation, we have designed VRLS to be applicable to road sections without network coverage, which can have arbitrary size and number of lanes, any number of vehicles, utilizing any resource pool configuration with different number of resources in time and frequency. VRLS consists of a deep neural network, which processes the state of the environment and decides on the resource allocation for vehicles as an outcome of its trained policy. The reward provided to the RL agent during its training is designed to maximize the reliability of V2V transmissions. In this paper, we extend the basic VLRS design with the following key contributions:}

\begin{enumerate}
	\item We train VRLS in simplified and simulated vehicular environments, and show that it can be deployed without further training in realistic, complex environments, varying in terms of mobility, wireless channel characteristics, OOC area size, network load, and traffic. 
	\item We evaluate the performance of VRLS in terms of reliability, user fairness, packet inter-reception time, and latency, as well as the impact of network quality of service on V2V applications, in comparison to the state-of-the-art distributed scheduler mode 4\footnote{Note that the specification of the successor standard NR mode 2 was not yet published during the preparation of this work.}. In terms of reliability, VRLS reduces by half the packet loss of mode 4 in highly loaded conditions, and performs close to the theoretical maximum in low-load scenarios. Further, VRLS does not compromise on fairness across the vehicular users, while achieving similar latency and higher ``mutual awareness'' (defined in Section \ref{KPIs}) as compared to mode 4.
	\item Considering that the network might need to operate differently configured resource pools {\color{black}in terms of the number of resources in time and frequency}, e.g., to support different V2V services, we show that VRLS can be trained across multiple predetermined resource configurations at once to support any of them by learning a single policy.
\end{enumerate}

In the rest of the paper, we first present the background and the related work in Section \ref{Sota}. Section~\ref{Model} defines our system model and the problem we address. We elaborate on our VRLS design in Section~\ref{Algorithm}. After describing our evaluation methodology in Section~\ref{Evaluation_Methodology}, we present our performance evaluation and analysis in Section~\ref{Results}, and the training performance of VRLS in Section~\ref{Curves}. Finally, Section~\ref{Conclusion} concludes the paper.

\section{{\color{black} Background and Related Work}}
\label{Sota}

{\color{black}We first present the challenges of V2V communications in terms of performance requirements in Section~\ref{ITS}. To satisfy these requirements, radio resource allocation plays a critical role, as we elaborate in Section~\ref{V2V_ResAlloc}. The centralized resource allocation mode introduced in cellular networks by 3GPP shows clear performance benefits over the distributed methods. We review the distributed resource allocation mechanism specified by 3GPP in Section~\ref{Mode4}, which we take as a baseline in our evaluations. Finally, we introduce the centralized, RL-based approach to resource allocation {\color{black}for V2V communications in areas delimited by cellular network coverage}, and identify its difference from the prior art in Section~\ref{Sota_RL}.}
	
\subsection{{\color{black} Intelligent Transport Systems and V2V Communications}}
\label{ITS}

V2V communications enable road safety and traffic efficiency via intelligent transport system (ITS) applications: autonomous driving and vehicle platooning, to name a few \cite{3gppTS22185}, \cite{3gppTS22186}. To achieve this, the V2V network has to satisfy the performance requirements of target use cases, most notably the reliability, latency, and throughput. The reliability requirements indicate the maximum tolerable packet loss rate between vehicles, together with a maximum application-level latency to be supported. LTE V2V use cases typically require $80-90$\% reliability with $100$~ms latency, targeting various basic road traffic safety applications \cite{3gppTS22185}, \cite{3gppTR36885}. On the other hand, advanced use cases to be supported by {\color{black}5G} NR, such as platooning vehicles with a high degree of automation, require up to $99.99$\% of reliability with $10$ ms latency \cite{3gppTS22186}. Such requirements assume an abundance of radio resources available for V2V communication. {\color{black}However,} under realistic conditions, where the availability of radio resources is limited, achieving high reliability becomes difficult. Under such conditions, resource allocation plays a vital role in the performance of V2V applications.

\subsection{{\color{black} Radio Resource Allocation for V2V Communications}}
\label{V2V_ResAlloc}
{\color{black}In the context of the 3GPP standard, V2V communications utilize a \textit{resource pool}, which consists of a limited {\color{black}number} of time-frequency resources configured by the network. Assignment of these resources to the vehicles in a reliable, timely, and efficient manner becomes a critical task in order to satisfy the demanding requirements of vehicular applications. 
	
Conventionally, the resource allocation task can be performed either in a centralized or distributed way.} {\color{black}The} authors in \cite{vukadinovic20183gpp} and \cite{nardini2018cellular} show that centralized resource allocation enables higher road traffic efficiency and safety, as compared to distributed scheduling mechanisms. To illustrate, the centralized scheduling {\color{black}mode} in LTE (called ``mode 3'' \cite{3gppTR37985}) can achieve a shorter inter-vehicle distance than LTE mode 4 and IEEE 802.11p in a high-density platoon with a guaranteed crash rate $\leq1\%$ \cite{vukadinovic20183gpp}. This is enabled by higher reliability and shorter latency of the V2V transmissions via efficient resource reuse that cannot be achieved by distributed scheduling. Despite providing very low latency, 802.11p is shown to suffer from the increasing collisions with the load{\color{black}\cite{vukadinovic20183gpp}}, and mode 4 is prone to collisions due to re-selection or persistent usage of the same resources by the vehicles in proximity{\color{black} \cite{bazzi2020blindspots}, \cite{yoon2021resolving}}. Irrespective of the underlying technology, {\color{black}the} reduced reliability degrades the performance of the V2V use cases (e.g., a platoon can either have a larger inter-vehicle spacing or can support fewer vehicles) \cite{giordano2019joint}.


\subsection{Distributed Scheduler LTE Mode 4}
\label{Mode4}

If {\color{black}the} centralized scheduling of V2V resources with the help of {\color{black}the} cellular network fails due to the loss of network connectivity, vehicles need to switch {\color{black}to the distributed resource allocation mode. Specifically,} upon experiencing a connection interruption long and often enough, vehicles stop using their network-scheduled resources and resort to a random resource selection procedure until the sensing results needed to operate LTE mode 4 become available \cite{3gppTR37985},{\color{black}\cite{3gppTS36300}}.

In mode 4, vehicles autonomously make semi-persistent resource (re-)selections from the configured resource pool for V2V communications, based on sensing{\color{black} \cite{3gppTS36213}, \cite{3gppTS36321}}. Specifically, as illustrated in Fig. \ref{Mode4Fig}, upon a message generation at time $t\textsubscript{gen}$, the vehicle selects a single resource from an upcoming resource selection window between $t\textsubscript{gen}+T_1$ and $t\textsubscript{gen}+T_2$. The vehicle can transmit using the selected resource on a periodic basis for $C\textsubscript{resel}$ times, i.e., semi-persistently, where $C\textsubscript{resel}$ takes a random value from a predefined interval. After $C\textsubscript{resel}$ transmissions, the vehicle makes a new resource re-selection with probability $1-P\textsubscript{keep}$, otherwise keeping the same resource by setting a new random value for $C\textsubscript{resel}$. Each resource (re-)selection is based on the sensing results of the past $1000$~ms from $t\textsubscript{gen}$, excluding the subframes the vehicle transmitted, {\color{black}as} no sensing was conducted due to the half-duplex radio constraint {\color{black}(vehicles can either transmit or receive, but not both at the same time)}. The vehicle further excludes the resources where it sensed an average received power larger than a predefined threshold $\text{Thr}\textsubscript{sense}$. It sorts the remaining resources with respect to their average Received Signal Strength Indicator (RSSI), and selects {\color{black}a} resource randomly from the top $20\%$ (the lowest RSSI). 

{\color{black}The} sensing mechanism enables vehicle{\color{black}s} to find {\color{black}a} ``free'' resource, or, in case of heavy resource use, a resource with less interference. On the other hand, {\color{black}the} randomization aims at mitigating {\color{black}the} persistent resource conflicts {\color{black}due to} multiple vehicles continuously selecting the same {\color{black}resource}. Nonetheless, since sensing measurements are limited in time and space, vehicles in mode 4 are prone to the well-known hidden-node problem. To illustrate, if vehicles far apart cannot sense each other's transmissions and select the same resource, their transmissions can interfere on a receiver located between them.

{\color{black} Several works have focused on improving the performance of mode 4. Most of them target {\color{black}the persistent resource collision problem}. The authors in \cite{jung2019persistent} and \cite{wendland2019mode4splitting} propose reserving resources at each resource selection instance and alternately using them to mitigate the collision probability. Other works propose exchange of information among the vehicles, such as channel measurements \cite{heo2021hv2xmode4}, location \cite{molina2019geobased}, or status and reservation information of resources \cite{cecchini2017maprp}, \cite{sabeeh2019estimation}, \cite{jeon2020explicit}. Revisions to the sensing mechanism are also proposed in several works, mainly by considering different weighting strategies for selecting the resources \cite{leon2018enhanced}, \cite{sharma2020context}. Overall, the proposed extensions to mode 4 either require additional signaling or increase resource occupancy, hence making them less efficient from resource utilization point of view. Further, the parameters of the mode 4 algorithm, as well as the extended methods, require careful tuning in order to achieve the desired performance.

\begin{figure}[!t]
	\centering
	\includegraphics[width=\columnwidth]{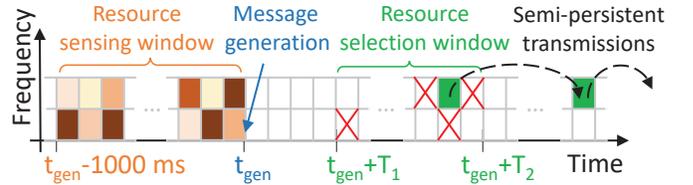}
	\vspace{-1.5\baselineskip}
	\caption{Illustration of the scheduling mechanism of LTE mode 4. Resources sensed with large power (represented with darker colors) are excluded from the selection (crossed). The selected resource (green) among the remaining ones is utilized to transmit semi-persistently.}
	\label{Mode4Fig}
	\vspace{-1\baselineskip}
\end{figure}

\subsection{{\color{black} RL-based Approach for V2V Radio Resource Allocation}}
\label{Sota_RL}

{\color{black} RL has recently gained significant attention in vehicular networks, with applications ranging from storage of data~\cite{wu2016reinforcement} to controlling of information flow in sensing networks~\cite{wang2017vehicular}, thanks to its ability to handle tasks in dynamic, time-varying environments. The resource allocation task challenges the mainstream approach of formulating an optimization problem and solving it (sub-)optimally, {\color{black}because of} the highly dynamic nature of vehicular networks~\cite{magazine}.} RL offers an alternative solution, which interacts with, and adapts its actions to the unknown environment. Further, sequential decision-making encountered in resource allocation tasks is a native functionality of RL\cite{liang2020deeplearning}. 

The majority of {\color{black}the} works applying RL to the resource allocation problem in vehicular networks consider joint optimization of V2V and cellular links \cite{ye2019deep}, \cite{hu2021deep}, \cite{yang2019intelligent}. The authors in \cite{xia2020cluster} utilize V2V links to cooperatively relay the cellular downlink data within a cluster of vehicles, with the aid of RL-based scheduling. RL is also applied to distributed resource allocation mechanisms \cite{pressas2017contention}, \cite{su2020coexistence}. While RL-based approaches are shown to solve resource allocation tasks efficiently, the cost of such solutions comes in terms of {\color{black}an} additional effort required for training prior to deployment. 

We have first applied RL for managing the V2V resources centrally in \cite{sahin2018reinforcement}, motivated by the above-mentioned works. We later proposed VRLS in \cite{sahin2019vrls}, where we demonstrated its ability to avoid {\color{black}resource conflicts}, and to reuse resources more efficiently as compared to the state-of-the-art V2V scheduling algorithms, in a basic OOC setting and under different resource pool configurations. Compared to {\color{black}the} previous works, our approach differs by proposing a \emph{centralized}, \emph{RL-based approach} for managing the resources for V2V communications \emph{outside the network coverage}, {\color{black}whereby resources are provided to the vehicles before they leave the coverage.}}

\section{System Model}
\label{Model}

\subsection{Vehicular Network Environment}
{\color{black} We consider a vehicular network where vehicles broadcast V2V messages with periodic and event-triggered, i.e., aperiodic, traffic. {\color{black} A typical example of periodic traffic is the regular broadcast of vehicle information such as position and speed, as in cooperative awareness messages (CAMs)~\cite{ETSIEN3026372}, whereas aperiodic traffic is triggered on events to warn vehicles such as of an accident, as in decentralized environmental notification messages (DENMs)~\cite{ETSIEN3026373}.} {\color{black}The V2V} messages are assumed to have a fixed size {\color{black}of} $S\textsubscript{msg}$, requiring a predefined {\color{black}number} of time-frequency resources based on the modulation and coding scheme (MCS). {\color{black}In {\color{black}the} case of {\color{black}the} periodic traffic, {\color{black}the} messages are generated with {\color{black}a} fixed periodicity {\color{black}of} $T\textsubscript{msg}$ by all vehicles. For {\color{black}the} event-triggered traffic, {\color{black}the} event arrivals are assumed to follow a Poisson process with an arrival rate of $X\textsubscript{evt}$ per s for each vehicle \cite{3gppTR36885}, upon which a single message is generated.\footnote{{\color{black}According to {\color{black}the} ETSI specifications \cite{ETSIEN3026372}, \cite{ETSIEN3026373}, while periodic by default, the periodicity of CAM transmissions can be adjusted depending on vehicular mobility and traffic load, and DENM transmissions might contain bursts of several periodic messages. We evaluate the impact of {\color{black}the} different modes of traffic separately, by following previous research work (e.g., \cite{bazzi2018study}, \cite{mansouri2019first}) and applying the traffic models proposed in 3GPP~\cite{3gppTR36885}. This approach makes the analysis easier and enables us to generalize our results to any type of V2V traffic beyond {\color{black}the} CAM and DENM applications.}}
			
We {\color{black} focus on} V2V communications in a delimited out-of-coverage area (DOCA). A DOCA can have any size and shape depending on the scenario (e.g., two-dimensional in urban and one-dimensional on highway). We assume a DOCA that consists of a two-way highway segment of length $L\textsubscript{DOCA}$ outside the coverage of BSs at its two ends, as illustrated in Fig.~\ref{VRLS}. Vehicles of length $L\textsubscript{veh}$ are assumed to travel on $J$ lanes per direction. {\color{black}The} BSs are assumed to be aware of the existence of the DOCA, as well as its location, length, etc.

\begin{figure}[!t]
	\centering
	\includegraphics[width=\columnwidth]{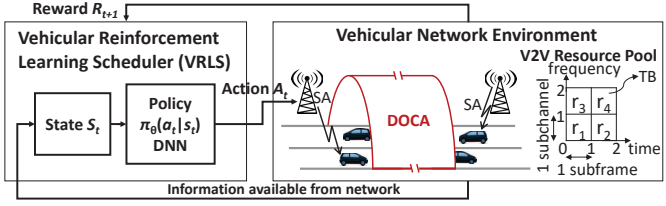}
	\vspace{-1\baselineskip}
	\caption{VRLS applied to our vehicular network environment, which is a delimited out-of-coverage area (DOCA). Vehicles communicate with each other in a DOCA using the resources indicated by the scheduling assignments (SAs) sent by the delimiting BSs before they enter the DOCA.} 
	\label{VRLS}
	\vspace{-0.5\baselineskip}
\end{figure}

\vspace*{-1\baselineskip}

\subsection{Time-Frequency Resources}
In line with the LTE and NR specifications, {\color{black} V2V communications outside the coverage utilize a dedicated pool of time-frequency resources configured by the network. We denote pool configuration by $C^{K \times M}$, where $K$ is the number of subchannels and $M$ is the number of subframes that the pool contains, over the frequency and the time domains, respectively, repeating with {\color{black}a} periodicity {\color{black}of} $T\textsubscript{msg}$.} Fig.~\ref{VRLS} illustrates an example resource pool configuration with $2$ subchannels and $2$ subframes, i.e., $C^{2 \times 2}$.

Following {\color{black}the} LTE V2V assumptions~\cite{3gppTR36885}, a transmission of a single V2V message occupies a single time-frequency resource $r \in C^{K \times M}$, which is referred to as a transmission block (TB)}. A TB consists of a single subframe of $1$~ms, and a single subchannel containing a sufficient number of resource blocks (RBs) {\color{black}to carry the message}. {\color{black}No redundant transmissions are considered. Consequently,} the latency of a {\color{black}successfully received} message is {\color{black}bounded} by the time-length of the pool, i.e., the number of subframes, to {\color{black}a} maximum of $M$~ms. 

We assume half-duplex radios, that is, vehicles transmitting at the same time are not able to receive each other's message (e.g., TBs $r_1$ and $r_3$ sharing the same subframe in Fig. \ref{VRLS}) \cite{ding2012combating}. We refer to {\color{black}the} unsuccessful reception of {\color{black}the} messages due to {\color{black}the} HD limitation as \textit{HD error} or \textit{conflict}, and the relation among the TBs in the same subframe causing this phenomenon, as \textit{HD constraint}, in the paper. When no HD errors occur, successful decoding of a message is further conditioned on the signal-to-interference-plus-noise ratio (SINR) measured for the corresponding TB at the receiver. {\color{black} The SINR of a single transmission at receiver $j$ from transmitter $i$ is:
\begin{equation}
	\text{SINR}_{ij} = \dfrac{P\textsubscript{Tx}|h_{ij}|^2}{\sigma^2+\sum^{L}_{l=1,l\neq i}P\textsubscript{Tx}|h_{lj}|^2},
\end{equation}
where $P\textsubscript{Tx}$ is the transmit power of the transmitter, and $|h_{ij}|^2$ denotes the channel coefficient between the transmitter $i$ and the receiver $j$, which accounts for the path loss and fading effects of the wireless channel on the transmitted signal. $\sigma^2$ is the noise power, and the summation term in the denominator denotes the interference due to {\color{black}the} other vehicles $l=1,...,L$ using the same TB as $i$.} The SINR depends on the interference level of {\color{black}the} other transmissions using the same TB, under the path loss and fading effects of the propagation channel. Messages transmitted by different vehicles using the same TB (i.e., the same subframe \textit{and} the same subchannel) may interfere, depending on the propagation conditions, and lead to decoding errors at the receiver, which we refer to as \textit{collision errors}. However, spatial reuse of the same TB or the TBs sharing the same subframe is possible among sufficiently far apart transmitters, without creating any collision and HD errors, respectively. Further, unsuccessful receptions could also result from the channel effects, e.g., due to path loss and shadowing that considerably reduce the received SINR (or SNR), which we refer to as \textit{propagation errors}.

\vspace*{-1\baselineskip}

\subsection{Problem Definition}
\label{ProblemDef}
In the absence of a cellular network assigning the resources for V2V communications, as in the case of a DOCA, vehicles can resort to distributed resource scheduling mechanisms based on sensing or using resources randomly \cite{3gppTR37985}. However, such distributed resource allocation methods result in degraded performance of V2V communications as compared to the centralized allocation of resources within the network coverage, given the same limited {\color{black}number} of resources available \cite{vukadinovic20183gpp}, \cite{bazzi2017}.

We argue that the areas without network coverage, i.e., DOCAs, are usually well known to the network (such as via map information providing the location of a tunnel or detecting link interruptions by measured signals at a certain location). Therefore, it is possible to create a centralized scheduler that ``pre-schedules'' the resources for the DOCA. The centralized scheduling entity is assumed to exchange information with the BSs delimiting the DOCA, where it can gather the information needed for scheduling. In turn, {\color{black}the} BSs forward the scheduling decisions to the vehicles via scheduling assignments (SAs) before the vehicles enter the DOCA. An SA indicates the resource to use from the next available resource pool period, every time the vehicle generates a message during its travel in the DOCA. Accordingly, the task is to assign a resource $r$ from the resource pool to each vehicle going into the DOCA, where the vehicle uses the assigned resource for all its V2V transmissions throughout its journey. 

{\color{black}Our objective is to find a centralized radio resource management algorithm that maximizes the reliability of {\color{black}the} V2V transmissions in the DOCA. To quantify the reliability, we use the packet reception ratio (PRR) metric as defined in the 3GPP standard~\cite{3gppTR36885}. {\color{black} For a single message transmitted from vehicle $i$, the PRR is calculated by $X_i/Y_i$, where $Y_i$ is the number of vehicles located within the range $(a, b)$ from the transmitter, and $X_i$ is the number of vehicles with successful reception among $Y_i$. The average PRR is then calculated for a series of messages consecutively transmitted by all the vehicles in the environment as $ (\sum^V_{i=1}\sum^{N_i}_{n=1}X_{i,n}) / (\sum^V_{i=1}\sum^{N_i}_{n=1}Y_{i,n}) $ with $N_i$ denoting the number of generated messages by vehicle $i$, and $V$ is the number of vehicles~\cite{3gppTR36885}. Assuming a resource pool configuration of $C^{K \times M}$, we would like to find a resource assignment $ (k,m) $ for each vehicle $i$ that maximizes the PRR of all vehicles, which can be formally expressed as:
		
\vspace{-0.5cm}
		
\begin{IEEEeqnarray}{rCl}
	\label{eqn_scheduler}
	&\argmax_{\{\alpha^i_{k,m}\}_{\forall i}} \sum^V_{i=1} \sum^{N_i}_{n=1} X_{i,n}, \; \text{where} \: X_{i,n}=\sum^{Y_{i,n}}_{j=1} \eta_{ij},& \\
	&\text{ s.t. }&\IEEEnonumber\\
	\IEEEyessubnumber
	& \eta_{ij} = \left\{ \,
		\begin{IEEEeqnarraybox}[][c]{l?s}
		1 & if $\dfrac{P\textsubscript{Tx}|h_{ij}|^2\alpha^i_{k,m}\beta^j_{m}}{\sigma^2+\sum^{V}_{l=1,l\neq i}P\textsubscript{Tx}|h_{lj}|^2\alpha^l_{k,m}} \geq \gamma$, \\
		0 & otherwise.
				\IEEEstrut
	\end{IEEEeqnarraybox}
	\right.
	\\ 
	\IEEEyessubnumber
	& \alpha^i_{k,m} = \left\{ \,
	\begin{IEEEeqnarraybox}[][c]{l?s}
		1 & if $i$ using subchannel $k$ and subframe $m$, \\
		0 & otherwise.
	\end{IEEEeqnarraybox}
	\right.
	\\
	\IEEEyessubnumber
	& \beta^j_m = \left\{ \,
	\begin{IEEEeqnarraybox}[][c]{l?s}
		1 & if $j$ using subframe $m$, \\
		0 & otherwise.
	\end{IEEEeqnarraybox}
	\right.
	\\
	\IEEEyessubnumber
	\label{eqn_scheduler_single}
	& \sum^M_{m=1} \sum^K_{k=1} \alpha^i_{k,m} = 1, \forall i.
	\\
	\IEEEyessubnumber
		\label{eqn_scheduler_power}
	& 0 < P\textsubscript{Tx} \leq P\textsubscript{max}
\end{IEEEeqnarray}

\vspace{-0.1cm}

In the above formulation, $ \eta_{ij} $ is a binary variable indicating whether the reception of a single message $n$ at receiver $j$, transmitted from $i$ is successful or not. Successful reception is conditioned on the $SINR_{ij}$ being above a target threshold $\gamma$, subject to collisions and the HD constraint represented by the following variables. $\alpha^i_{k,m}$ is a binary variable denoting whether vehicle $i$ is transmitting using the subchannel $k$ and subframe $m$, and $ \beta^j_m $ imposes the HD constraint on vehicle $j$ if it is transmitting on the same subframe $m$ that it was supposed to receive another transmission. Constraint~(\ref{eqn_scheduler_single}) indicates that each vehicle transmits at most using a single TB in a given period of the resource pool. Note that the resource assignments to different vehicles are not necessarily unique, i.e., the same resource could be reused by different vehicles. Finally, constraint~(\ref{eqn_scheduler_power}) limits the transmit power of vehicles.}

{\color{black}Solving the above optimization problem is a challenging task given the limited number of resources to be managed for tackling the HD and collision constraints in a highly dynamic environment beyond the network coverage.} As a solution, we propose an RL-based scheduling approach called VRLS. VRLS is a logical entity, deployed at the network infrastructure, whose function is to allocate resources to vehicles entering the DOCA with the goal of maximizing the reliability of their V2V transmissions in the DOCA. While it is trained for maximizing the PRR in the DOCA, which can be conducted off-line, in operation, VRLS solely utilizes the information already available at the network.}

\section{VRLS: Vehicular Reinforcement Learning Scheduler}
\label{Algorithm}

 \begin{figure*}[ht!]
	\begin{center}
		\includegraphics[width=2\columnwidth]{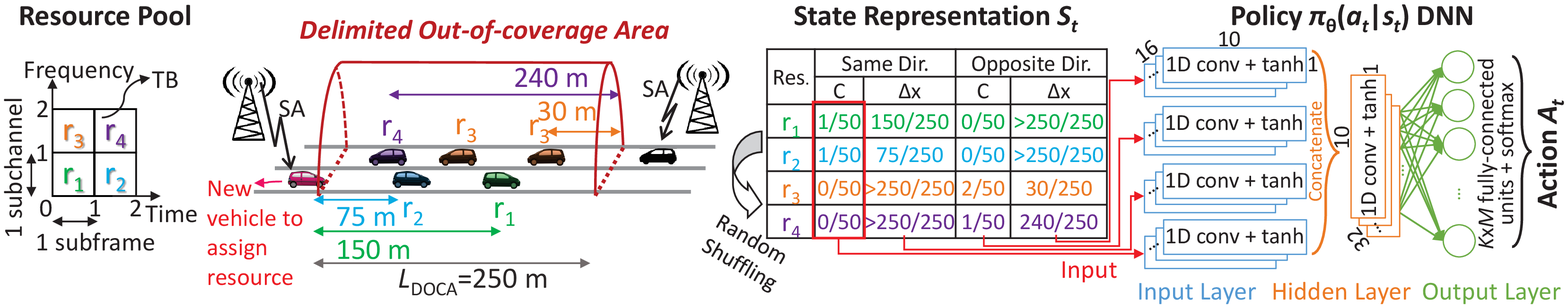}  
	\end{center}
	\vspace{-1.5\baselineskip}
	\caption{State representation of a simple exemplary scenario in the DOCA provided to the DNN of VRLS.}
	\label{VRLSdesign}
	\vspace{-1\baselineskip}
\end{figure*}

We formulate the centralized resource pre-allocation problem for the DOCA as a single-agent RL problem, where VRLS acts as the agent on the vehicular network environment. Based on the observed \textit{state} $S_t$ of the environment at each discrete instant $t$ in which a new vehicle arrives at the DOCA, VRLS takes an \textit{action} $A_t$, which is to assign a single time-frequency resource to that vehicle. The actions of VRLS are based on its trained \textit{policy} $\pi$, which we model as a \textit{deep neural network} (DNN). The agent is trained with a \textit{reward} signal $R_{t+1}$ provided upon each action, indicating how “good” the action was. In turn, the training goal of the agent is to maximize the total reward it receives in the long run. We train VRLS using the state-of-the-art \textit{actor-critic} RL algorithm.

Typically, RL solutions in the literature are designed, trained, and evaluated in the same environment that has a specific distribution of (or even fixed) parameters. However, particular design choices tailored for a specific setting may not work as well or even be applicable when the parameters of the environment change significantly \cite{gamrian2018transfer}. We have observed the same problem in our earlier work \cite{sahin2018reinforcement}, where we had to re-design the components of our solution (thus, also re-train) in order to target different vehicular environments. This turns into a serious limitation, considering the diverse environments that V2V communications need to support. Considering that the RL agent is trained ``off-line'', i.e., before its deployment, it is desirable to learn a policy that is applicable to different environments of interest without further training. This would eliminate the need of training a new agent from scratch every time an unseen (yet similar) condition arises in the deployed environment. Furthermore, it would offer the possibility to train the agent in a simpler, simulated environment, saving from the burdens of real-world training.

As we elaborate in the following, we design the state, action, reward, and training of VRLS to be applicable to any DOCA environment having an arbitrary size and number of lanes, with any number of vehicles inside, utilizing any resource pool configuration with an arbitrary number of resources in time and frequency. These design features aim at making VRLS a practical solution that is deployable in a variety of realistic OOC environments with different mobility, network load, and wireless conditions, and further facilitate efficient and practical training over simpler and simulated ones.

\vspace{-0.2cm}
\subsection{State Representation}
\label{State}
We devise the state $S_t$ to provide the agent with information on \textit{how} the resources are utilized at each instant $t$ a vehicle is entering the DOCA. Formally, $S_t$ is a matrix (shown in Fig.~\ref{VRLSdesign}) with each row representing a resource (TB) in the resource pool, and the columns providing the following information:

\begin{itemize}
	\item $C$: number of vehicles each resource is assigned, normalized to the maximum number of vehicles {\color{black}that the DOCA can accommodate} (derived by {\color{black} $J \times L\textsubscript{DOCA} / L\textsubscript{veh} $}, considering the case where all lanes are fully occupied). $C$ represents whether the resources are free and, if not, how much loaded. The vehicle density in the DOCA could also be obtained from $C$ by accounting for the sum of the allocated resources in proportion to the calculated maximum number of vehicles.
	\item  $\Delta x$: distance from the entrance point of the DOCA to the latest vehicle the resource was assigned to, normalized to $L\textsubscript{DOCA}$. The distance is estimated by multiplying the amount of time passed since the vehicle went out of coverage by the average speed $v\textsubscript{avg}$ of the vehicles in the DOCA, as their speed might vary over time. $\Delta x$ represents how far the potential interferers are, hence facilitating spatial reuse of each resource.
	\item The \textit{order} of the columns represents the direction of the vehicle entering the DOCA. The first pair of columns provides $C$ and $\Delta x$ for the vehicles traveling in the same direction as the vehicle entering the DOCA, while the second pair provides the information from the opposite direction of the DOCA.
\end{itemize}

Algorithm~\ref{algo_state} details how $S_t$ is calculated. The following variables are input for each vehicle $i$ inside the DOCA: time of its entry $t_i$, assigned resource $r_i$, traveling direction (i.e., east or west) $d_i$; as well as the average speed of the vehicle traffic $v\textsubscript{avg}$ and the current time $t\textsubscript{now}$. 

$S_t$ is applicable to any number of resources and vehicles, and any DOCA size, thanks to the normalized state variables. An example $S_t$ is illustrated in Fig.~\ref{VRLSdesign} for a simple scenario with $4$ resources and a DOCA of {\color{black}$L\textsubscript{DOCA}=250$~m}, where a maximum of $50$ vehicles {of \color{black}$L\textsubscript{veh}=5$~m} can fit per lane.

\vspace{-0.2cm}
\subsection{Action Definition}
\label{Action}
The agent takes an action $A_t$ at each instant $t$ {\color{black} a vehicle is about to enter} the DOCA. In case {\color{black} multiple vehicles} enter at the same instant, the corresponding actions are taken in random order. Action {\color{black}denotes assigning} a single time-frequency resource, i.e., a TB, which the vehicle uses for its V2V transmissions through the DOCA. {\color{black} Accordingly, the action-space is a vector of $K\times M$ TBs} in the resource pool configured in the network. VRLS gives the decision on which TB to be assigned at time $t$ by its policy $\pi$. The policy is a mapping $\pi(a_t | s_t) \rightarrow[0,1]_{K\times M}$ from the state $S_t$ of the environment at $t$, to a probability distribution over the set of possible actions (the TBs in the resource pool). The TB to be assigned is selected at random according to this distribution. 

\begin{algorithm}
	{\color{black}
		\caption{Calculation of the state representation $S_t$}
		\label{algo_state}
		\begin{algorithmic}[1]
			\Require $t_i, r_i, d_i \; \forall i, v\textsubscript{avg}, t\textsubscript{now}$ 
			\State Update vehicle distances: $\Delta x_i \gets v\textsubscript{avg} (t\textsubscript{now}-t_i) \; \forall i$
			\State Remove vehicle if it left the DOCA ($\Delta x_i > L\textsubscript{DOCA}$)
			\For{each resource in the pool $r=1,2,...,K\times M$}
			\For{each road direction $d=\{\text{east},\text{west}\}$}
			\State Find vehicles using the same resource in the same direction (check if $d_i == d \; \&\& \; r_i == r$)
			\State Update $C$ of the respective $d$ and $r$ with the number of found vehicles 
			\State Sort distances of found vehicles to find $\min(\Delta x_i)$
			\State Update $\Delta x$ of respective $d$ and $r$ with $\min(\Delta x_i)$
			\EndFor
			\EndFor
			\State Order columns of $S_t$ w.r.t. direction of the entering vehicle\\
			\Return $S_t$
		\end{algorithmic}
	}
\vspace{-0.1cm}
\end{algorithm}

\vspace{-0.2cm}
\subsection{Deep Neural Network Architecture}
\label{DNNs}
{\color{black} The large space of possible combinations of vehicles and resources makes tabular RL methods infeasible for this problem \cite{sutton1998reinforcement}. This leads us to apply approximate solution methods by utilizing a DNN to represent the policy. DNN {\color{black}consists of a set of adjustable parameters $\theta$, i.e., $\pi_\theta(a_t | s_t)$ that maps} a given state to action probabilities.} 

We utilize a convolutional neural network (CNN) to model $\pi_\theta$. {\color{black}At the input layer, we utilize $ 4 $ sets of convolutional filters, each processing a different column of $S_t$, as illustrated in Fig.~\ref{VRLSdesign}. Each set contains $16$ 1D convolutional filters of length $10$ and applies a $\tanh$ nonlinearity. The output of these filters is then concatenated and input to the hidden layer of the CNN, which is another convolutional layer with $32$ 1D filters of length $10$. The output layer of the CNN is a fully-connected layer with the number of units equal to the number of actions, i.e., $K\times M$ TBs available in the configured resource pool. The $\operatorname{softmax}$ activation function is applied at the output to produce a probability distribution over the actions, from which the TB to be assigned is selected at random.}

\vspace{-0.2cm}
\subsection{Data Augmentation}

The output of convolutional layers is variant to the \textit{order} of the input data they process, due to the convolution operation. Although this is useful for their most common applications, such as processing images or audio that present naturally ordered data (e.g., ordered pixels in space), this feature poses a limitation in our case. The policy of VRLS should not depend on the order of resources presented in $S_t$, but rather on the information provided about them. That said, the HD constraint depends on the order of resources in the resource pool, as the HD error is caused by using the resources in the same subframe. For example, the two pool configurations $C^{4 \times 5}$ and $C^{2 \times 10}$ have different HD constraints, although both have the same number of resources, and there is no information in the state representation to differentiate between them.

To address this challenge, we resort to data augmentation methods. Data augmentation is commonly utilized in deep learning, e.g., for image classification tasks, where the agent is made to \textit{learn} becoming invariant to modifications of the data, e.g., image rotation, clipping, etc., by providing the agent with such modified inputs during the training \cite{imagenet}. In our case, we apply data augmentation by randomly shuffling the order of resources first in time and then in frequency. The rows of $S_t$ and the resource selection probabilities at the output layer of the DNN follow this order. {\color{black}To illustrate our method, consider the example in Fig.~\ref{VRLSdesign}. A raster-scan ordering of the resources in the resource pool is ${[r_1,r_2,r_3,r_4]}$. We first group the resources sharing each subframe (corresponding to ``columns'' of the resource pool), and randomize the order of these groups. This yields a raster-scan ordering of, e.g., ${[r_2,r_1,r_4,r_3]}$. Then, we group the resources sharing each subchannel, i.e., the ``rows'' of the pool, and randomly shuffle the order of the ``rows''.} This way, the convolutional network becomes invariant to the order of resources in time or frequency, while being able to infer the HD constraints among the resources.

\subsection{Reward Definition}
\label{Reward}
We incorporate the reliability metric PRR into the reward signal $R_{t+1}$ as a linear function of it: $R_{t+1}=-10\times(1-$PRR$)$. PRR is computed at a certain range of interest for all transmissions within the DOCA \textit{since the last action}, i.e., in between each vehicle arrival to the DOCA. {\color{black}The range at which the PRR is measured for the reward could be determined by several factors, such as the distance at which a target PRR value needs to be satisfied; additionally, it can also be limited by the transmission power of the vehicles.} In case no transmissions take place between consequent actions, e.g., when two vehicles enter the DOCA almost at the same time, we provide the reward of the previous action to the agent.} 

\begin{table*}[]
	\renewcommand{\arraystretch}{1.1} 
	\centering
	{\color{black}
	\caption{Algorithmic Complexity of VRLS}
	\vspace{-0.8\baselineskip}
	\label{table_complexity}
			\centering
			\begin{tabular}{|r|l|l|}
				\hline
				\multicolumn{1}{|l|}{} & \multicolumn{1}{c|}{\textbf{Time complexity}} & \multicolumn{1}{c|}{\textbf{Space complexity}} \\ \hline
				\textbf{\begin{tabular}[c]{@{}r@{}}Stage (i): Calculation of $ S_t $\\ (Steps of Algorithm~\ref{algo_state})\end{tabular}} &
				\begin{tabular}[c]{@{}l@{}}Step 1: $\mathcal{O}(V)$\\ Step 2: $\mathcal{O}(V)$\\ Step 5: $\mathcal{O}(KMV)$\\ Step 7: $\mathcal{O}(KMV\log V)$\\ Overall: $\mathcal{O}(KMV\log V)$\end{tabular} &
				\begin{tabular}[c]{@{}l@{}}$3V+2+4KM$ variables \\ ($t_i, r_i, d_i \; \forall i=1,...,V$, $v\textsubscript{avg}$, $t\textsubscript{now}$, and $S_t$) \\ Overall: $\mathcal{O}(V+KM)$\end{tabular} \\ \hline
				\textbf{Stage (ii): CNN processing} &
				\begin{tabular}[c]{@{}l@{}}Input layer: $\mathcal{O}(K M)$\\ Hidden layer: $\mathcal{O}(K M)$\\ Output layer: $\mathcal{O}(K^2 M^2)$\\ Overall: $\mathcal{O}(K^2 M^2) $\end{tabular} &
				\begin{tabular}[c]{@{}l@{}}Input layer: $4 \times 16 \times 10$ variables\\ Hidden layer: $32 \times 10$ variables\\ Output layer: $32\times(4 \times 16 (KM – 10 + 1) – 10 + 1) \times KM$ variables\\ Overall: $\mathcal{O}(K^2M^2)$\end{tabular} \\ \hline
				\textbf{Total}         & $\mathcal{O}(K^2 M^2 + KMV\log V) $           & $\mathcal{O}(K^2M^2+KM+V)$                            \\ \hline
				\multicolumn{3}{c}{\vspace{-0.2cm}}  \\  
				\multicolumn{3}{c}{\begin{tabular}[c]{@{}c@{}}$ V $: number of vehicles within the DOCA; $ K $, $ M $: number of subchannels and subframes of the resource pool configured for V2V communications. \end{tabular}}
			\end{tabular}
	}
\vspace{-1.5\baselineskip}
\end{table*}

\subsection{Training Algorithm}
\label{TrainingAlgorithm}

In any RL task, the goal of the agent is to maximize its \textit{return} $G_t$, i.e., the future rewards it receives in the long run. In its simplest form, $G_t=\sum_{l=1}^{L_\text{epoch}} R_{t+l}$, given an \textit{epoch} of experience that consists of $L_\text{epoch}$ sequential tuples of state-action-reward. The return is estimated via a \textit{value function} $v(s_t)$, indicating how ``good'' it is for the agent to be in a state and to follow the policy $\pi$ onwards. Given the large state space of our problem, we also parametrize the value function with a DNN, with a set of parameters $w$. $v_w(s_t)$ utilizes the same architecture as the policy DNN in Section \ref{DNNs}, except for its final layer, which is a single fully connected neural unit that outputs the value of the given state.

We employ a state-of-the-art RL algorithm called \textit{policy-gradient actor-critic} \cite{sutton1998reinforcement} to train both DNNs. ``Actor'' refers to the component that learns the parameterized policy $\pi_\theta(a_t | s_t)$, and ``critic'' is the component that learns the parameterized value function $v_w(s_t)$ in order to evaluate the actor's policy, i.e., to ``criticize'' it. The actor updates the policy parameters $\theta$ by applying gradient ascent (with step-size $\alpha$) in the direction the critic indicates {\color{black}as} ``good''. The critic evaluates the actor's policy by measuring $\delta=G_t-v_w(s_t)$, i.e., the error between $v_w(s_t)$ that it estimated and the actual return received from the environment. The parameters $w$ of the critic are also trained in the direction indicated by $\alpha\times\delta$.

To improve training efficiency, we utilize parallel training. Specifically, we employ multiple learning actors, referred to as \textit{workers}, as proposed in \cite{Mnih2016}. A total of $N_\text{worker}$ workers are executed in multiple instances of the training environment in parallel. Each instance is simulated with a different random seed and initialized with a random resource assignment to the vehicles, with the first action being also randomized. Each worker asynchronously gathers epochs of experience (of length $L_\text{epoch}$) in its own environment. After every epoch, the worker updates the parameters of a {\color{black}single} global policy and a value function that are shared by all the workers. In the end, we obtain {\color{black}the global policy} that is trained using multiple workers, to be deployed in the target environment. This approach is beneficial in terms of speeding up the training thanks to a more efficient exploration of the state-space of the environment under distinct policies of multiple actors, {\color{black}and the added diversity helps to learn a general policy applicable to similar environments \cite{Mnih2016}.}

\vspace{-0.4cm}
\subsection{{\color{black}On Real-world Implementation of VRLS}}
\label{RealWorld}

{\color{black}In this study, we train and evaluate the performance of VRLS in simulative environments. Yet, the proposed methods might as well be implemented in a real-world vehicular network. In a real-world scenario, network vendors or operators would implement VRLS as an intelligent controller deployed at the edge of the network and integrated into the radio access network (RAN), thanks to the enabling architecture envisioned for 5G and beyond networks \cite{bonati2020open}. Within this architecture, the BSs deployed at the entrance/exit of the DOCA can be realized as remote radio units. While these radio units serve physical layer functions, they are connected to a centralized entity that is responsible for resource allocation and other higher-layer functionalities, where VRLS can be implemented. By implementing VRLS, operators would aim at ensuring seamless quality of V2V communications when vehicular users experience coverage losses. This would in turn ensure safer and more efficient road traffic. {\color{black}With regards to deployment and operation costs of VRLS, since being a logical entity, it can be implemented as software and can make use of the processing hardware available at the network infrastructure.} Additional processing power is necessary to train the RL agent (with high processing requirements and the possible need to pre-train on the simulated environments), and to operate it (with relatively low processing requirements). VRLS can again benefit from the fact that in 5G RAN, there are more computational resources deployed at the network edge to train learning algorithms \cite{bonati2020open}.
	
To train and operate VRLS, the BSs delimiting the DOCA would collect and report the required data constituting the state information input to VRLS as described in Section~\ref{State}. The BSs can easily keep track of the time of entry $t_i$ and assigned resource $r_i$ for each vehicle, autonomously, thus not requiring any additional signaling between the vehicles and the BS. Further, the BSs can obtain the information pertaining to $d_i$ and $v\textsubscript{avg}$ from the regular V2V traffic such as CAMs that vehicles transmit, thus again not requiring any additional signaling. The collected information at the BSs is forwarded to the centralized agent when an action is required. In turn, the actions of VRLS, namely the resource allocation, will be signaled to the vehicles via the BSs, before they enter the DOCA from the respective direction. {\color{black}Considering a pool configuration of $C^{K \times M}$, signaling of an assigned resource would consist of $\log(K+M)$ bits of information. Assuming a vehicle traffic of $0.4$ vehicles/s/lane arriving at a DOCA with $3$ lanes per direction \cite{3gppTR36885}, and a pool of $C^{2 \times 100}$, this would correspond to $ \sim 8 $ bits/s of downlink data traffic per BS.}

In Table~\ref{table_complexity}, we provide the algorithmic complexity of VRLS during its real-time operation (i.e., online inference phase), by decomposing it into two stages: i) the calculation of $S_t$ as described in Algorithm~\ref{algo_state}; and ii) the processing of $S_t$ by the trained CNN to select the resource as described in Section~\ref{DNNs}. Stage (i) yields a time complexity of $\mathcal{O}(KMV\log V)$, which is due to Step 7 that sorts at most $V$ vehicles in the DOCA, for each resource and direction, i.e., $2KM$ times. At stage (ii), the time complexity of the CNN is dominated by the multiplications at its fully-connected output layer, which yields $\mathcal{O}(K^2 M^2)$. Altogether, when both stages are combined, VRLS has a time complexity of $\mathcal{O}(K^2 M^2 + KMV\log V) $ during its real-time operation\footnote{{\color{black}We provide the precise details of the time complexity analysis in an online appendix accessible via https://doi.org/10.6084/m9.figshare.19365056.v1}}, thereby allowing a practical implementation. 

In terms of memory requirements, the algorithm in stage (i) stores $3V+2+4KM$ variables ($3$ per vehicle, $v\textsubscript{avg}$, $t\textsubscript{now}$, and $S_t$ having $4KM$ entries), which results in a space complexity of $\mathcal{O}(V+KM)$. At stage (ii), CNN stores a total of $4 \times 16 \times 10$, $32 \times 10$, and $32\times(4 \times 16 (KM – 10 + 1) – 10 + 1) \times KM$ parameters at its input, hidden, and output layer, respectively, hence yielding a space complexity of $\mathcal{O}(K^2M^2)$. Overall, the space complexity of VRLS is $\mathcal{O}(K^2M^2+KM+V)$, which is practical from the implementation point of view.

For training VRLS, it is possible to collect the reward signal from the network also in a real-world implementation. For example, vehicles could keep track of sent/received V2V message IDs with time and location stamps, which they report to the BSs after going back to the coverage. In turn, the BS calculates the PRR using this information to derive the reward. {\color{black}Such report sent by each vehicle would consist of the IDs and time/location stamps of the messages it has transmitted and received during its past travel within the DOCA.} {\color{black}We illustrate the incurred overhead with an example setting as follows. Assuming an average vehicle speed of $ 50 $ km/h with an arrival rate of $ 0.4 $ vehicles/s/lane, there will be $173$ vehicles in a DOCA of $ 1000 $ m with $3$ lanes/direction at a given time, on average. It would take $72$~s on average for a vehicle to travel through the DOCA, where it transmits $ 720 $ V2V messages, and receives at most $ 123840 $ messages from other vehicles, assuming a message transmission rate of $ 10 $ MHz (i.e., $T\textsubscript{msg}=100$ ms~\cite{3gppTR36885}) per vehicle, and all transmissions being successfully received by all vehicles. Further assuming that vehicle IDs are represented with $10$ bits of information, and it takes $16$ bits to represent the timestamp \cite{ETSIEN3026372} and $64$ bits to represent the location stamp of each message \cite{ETSIEN1028942}, each vehicle would then collect and report $1.401$ MB of information to the BS. This would correspond to around $1.69$ MB/s of uplink traffic per BS on average.} The delay in gathering the information does not pose a limitation for training since the agent acquires experience (the sequence of state-action-reward tuples) in batches before each training step.

On the other hand, training VRLS in simulative environments (and, if needed, re-training during a real-world usage) would circumvent numerous challenges associated with real-world training from scratch. By simulation, it is easier to create and collect sufficient data; hence the training becomes more flexible and less time-consuming. Besides, the costs of additional signaling and processing overhead at the network and at the vehicles required to collect data would be avoided.}
\vspace{-0.5\baselineskip}

\section{Evaluation Methodology}
\label{Evaluation_Methodology}

\subsection{Key Evaluation Metrics}
\label{KPIs}
{\color{black} In addition to the reliability metric PRR defined in Section~\ref{ProblemDef}, we also consider the following performance metrics:}

\begin{itemize}
	\item Mutual awareness: We use the mutual awareness metric \cite{mate16}, \cite{An2011}  to study the impact of PRR on the performance of applications running over V2V links. The authors in \cite{An2011} propose \textit{awareness probability} $P\textsubscript{A}$ as an intermediate metric that relates {\color{black}the} network quality of service {\color{black}to the} application performance. $P\textsubscript{A}$ is defined as ``probability of successfully receiving at least $n$ packets from a transmitter within the application tolerance time window $T$'', i.e., $P\textsubscript{A}=\sum_{n}^{k}\binom{k}{n}p^n(1-p)^{k-n}$, where $p$ is the PRR at the transmitter-receiver range of interest, and $k$ is the number of packets sent during $T$ \cite{An2011}. {\color{black} Thus, $P\textsubscript{A}$ reflects the network performance in the form of PRR, i.e., reliability, and can be used to evaluate its impact on the performance of V2V applications. Each application can set requirements on $P\textsubscript{A}$, as well as on $n$ and $T$. Requirements of several V2V applications are exemplified in \cite{An2011}, which we provide in Section \ref{NetworkToApp}.}
		
	\item Fairness: Since the PRR is reported {\color{black}for all vehicles and} all V2V transmissions, this does not {\color{black}indicate} whether all vehicles experience the same (or similar) reliability. Therefore, the following fairness metric {\color{black}is} additionally used in this paper: the average $ \text{PRR}_{j} $ {\color{black}is} computed separately for each vehicle $ j $, and afterward, the standard deviation of these per-user averages {\color{black}is} estimated. 
	
	\item Packet inter-reception time (PIR): {\color{black}PIR} is defined as the time elapsed between two successive successful receptions at a certain vehicle, transmitted from another one \cite{3gppTR36885}. It is used to evaluate the “situational awareness” of the vehicles at the V2V application layer \cite{vanet_pir}. 
	
	\item Latency: An important scheduling metric is the latency of V2V messages. The latency is measured between the time a V2V message is generated at the transmitter and successfully received by the receiver, at their application layers, respectively.  
\end{itemize}

\begin{table*}[h!]
	\renewcommand{\arraystretch}{1.1} 
	\setlength{\tabcolsep}{5pt} 
	\caption{Simulation Parameters}
	\vspace{-1\baselineskip}
	\label{Parameters}
	\resizebox{2\columnwidth}{!}{%
		\centering
		\begin{tabular}{|l|c|c|c|c|c|}
			\hline
			\multirow{2}{*}{ } & \textbf{Training Environment} & \multicolumn{4}{c|}{\textbf{Evaluation Environments}}\\
			\cline{2-6}
			& \textbf{E0} & \textbf{E1-L} & \textbf{E1-HL} & \textbf{E2-L} & \textbf{E2-HL}\\
			\hline
			
			\multirow{1}{*}{Mobility model} &  \multicolumn{5}{l|}{Straight highway section with $4$-m lane-width; vehicle length of $L\textsubscript{veh}=5$~m} \\
			\cline{2-6}
			\multirow{2}{*}{-- Highway scenario} & $J=1$~lane/direction; no overtaking & \multicolumn{2}{c|}{$J=1$~lane/direction; no overtaking} & \multicolumn{2}{c|}{$J=2$ lanes/direction with overtaking} \\
			& $L\textsubscript{DOCA}=500$~m & \multicolumn{2}{c|}{$L\textsubscript{DOCA}=\{500, 1000\}$~m} & \multicolumn{2}{c|}{$L\textsubscript{DOCA}=\{500, 1000\}$~m}\\
			\cline{2-6}
			
			-- Vehicle speeds & $50$~km/h (constant) & $\sim\mathcal{N}(120,12)$ km/h & $\sim\mathcal{N}(50,5)$ km/h & $\sim\mathcal{N}(120,36)$ km/h & $\sim\mathcal{N}(50,15)$ km/h\\
			
			\cline{2-6}
			\multirow{2}{*}{-- Dynamics} & $30$ vehicles with wrap-& \multicolumn{4}{c|}{Poisson arrival per direction with $\sim \operatorname {Exp}(0.4)$ \cite{3gppTR36885}}\\
			& around $\sim \operatorname {Exp}(0.4)$ \cite{3gppTR36885}& \multicolumn{4}{c|}{Realistic SUMO mobility \cite{sumo2018}}\\
			
			\hline

			\multirow{2}{*}{Network model} & TB interference/HD model & \multicolumn{4}{c|}{Complete LTE V2X protocol stack in ns-3 \cite{ns3} \cite{ns3D2D} \cite{ns3LTE}}\\
			& 						  & \multicolumn{4}{c|}{Bandwidth = $10$ MHz ($50$ RBs) with $32$ RBs active; Carrier frequency = $5.9$ GHz}\\
			&  						  & \multicolumn{4}{c|}{1 subchannel = $16$ RBs, 1 subframe = $1$ ms, MCS index = $9$}\\								   
			
			\hline
			
			\multirow{2}{*}{V2V channel model} & Protocol model \cite{gupta00} with  & \multicolumn{4}{c|}{3GPP channel model \cite{3gppTR36885} with path loss: LOS model in WINNER+B1 with}\\
			& Tx range $R\textsubscript{Tx}=120$ m & \multicolumn{4}{c|}{antenna height = $1.5$ m; path loss at $3$ m is used for distances $<3$ m}\\
			& No path loss and shadowing & \multicolumn{4}{c|}{Shadowing fading: log-normal distr. with $3$ dB std. dev. and $25$ m decorr. distance}\\
			&    & \multicolumn{4}{c|}{$P\textsubscript{Tx}=\{-5,23\}$ dBm; thermal noise level = $-174$ dBm/Hz}\\
			&	 & \multicolumn{4}{c|}{1 Tx and 2 Rx omni-directional antennae with $3$ dBi gain and $9$ dB Rx noise figure}\\
			\hline
			V2V message traffic model & \multicolumn{5}{l|}{$S\textsubscript{msg}=190$ B; $T\textsubscript{msg}=100$~ms for the periodic \cite{3gppTR36885}, and $X\textsubscript{evt}=1/$s for the aperiodic traffic.}\\ 
			\hline
			V2V resource pool	   & \multicolumn{5}{l|}{$C^{2 \times 10}$ ($2$ subchannels by $10$ subframes) and $C^{2 \times 50}$ ($2$ subchannels by $50$ subframes), periodically repeating with $100$~ms}\\
			\hline
			
			\multicolumn{6}{|c|}{\textbf{Mode 4 Configuration Parameters}}\\
			\hline
			\multicolumn{6}{|l|}{$T_1 = 4$ ms \cite{3gppTR36885}, $T_2=\{14,54\}$ ms, $C\textsubscript{resel}\sim \operatorname {Unif}[5,15]$ \cite{3gppTS36321}, $P\textsubscript{keep}=0$, $\text{Thr}\textsubscript{sense}=-120$ dBm}\\
			
			\hline
			\multicolumn{6}{|c|}{\textbf{VRLS Training Parameters}}\\
			\hline
			\multicolumn{6}{|l|}{$N_\text{worker}=16$; $L_\text{epoch}=60$; $\alpha=10^{-3} / (1 + 0.01\times \#ep^{1.1})$}\\
			\hline	
		\end{tabular}
	}
	\vspace{-1\baselineskip}
\end{table*}

\subsection{Training Environment Model and Methodology}

The training environment (denoted as “E0”) has basic vehicular mobility and wireless channel characteristics, which enables an efficient training thanks to reduced simulation time. The communications in the training environment is abstracted by the \textit{protocol model} \cite{gupta00}, i.e., a transmission is assumed to be successful if{\color{black}: i) }no other transmitter is using the same TB within a transmission range of {\color{black}$R\textsubscript{Tx}=120$~m} from the receiver{\color{black}; and ii) the receiver is not transmitting at that time (HD constraint)}. The impact of path loss and fading in the training environment is simplified by assuming correct decoding of the received packets {\color{black}within a range of} $R\textsubscript{Tx}$, and unsuccessful reception beyond that distance. {\color{black}The} mobility is simple with {\color{black}$30$} vehicles having the same constant speed of {\color{black}$50$~km/h, initially placed uniformly at random inside a DOCA of length $L\textsubscript{DOCA}=500$~m with $J=1$ lane/direction of $4$~m width}. {\color{black}Upon exiting the DOCA}, the vehicles are returned back from the opposite direction after a time offset $\sim \operatorname {Exp}(0.4)$, leading to an average inter-vehicle gap of $2.5$~s \cite{3gppTR36885}. The V2V resource pool in the network is assumed to be configured {\color{black}with $C^{2 \times 10}$, i.e.,} $2$ subchannels by $10$ subframes, {\color{black} to generate loaded conditions} with a V2V message traffic that has a fixed periodicity of $T\textsubscript{msg}=100$~ms.

The training is conducted using {\color{black}$N_\text{worker}=16$} workers in parallel, each interacting with a different instance of the environment in epochs of length $L_\text{epoch}=60$. Considering the transmission range of $R\textsubscript{Tx}=120$~m and the usage of the protocol model, PRR is measured at a range {\color{black}of} $0-100$~m away from the transmitters when calculating the reward signal $R_{t+1}$. {\color{black} Such {\color{black}a} range is crucial to avoid imminent crashes between vehicles, where communication reliability needs to be ensured according to V2V service requirements \cite{3gppTR22885}.}

\subsection{Evaluation Environment Models and Methodology}
We evaluate the performance of VRLS trained on the simple environment E0, over realistic environments accommodating various mobility, density, wireless channel conditions, and message traffic in the DOCA. To simulate the vehicular network in the evaluation environments, we use a full-stack LTE V2V protocol in ns-3 \cite{ns3}, based on our own implementation that extends the openly available D2D \cite{ns3D2D} and LTE \cite{ns3LTE} modules. The V2V channel model consists of a realistic path loss and shadow fading according to the 3GPP evaluation methodology \cite{3gppTR36885} (cf. Table \ref{Parameters} for details). Further, we simulate the vehicular mobility in the evaluation environments using the realistic traffic simulator SUMO~\cite{sumo2018}. Mobility traces from the simulated scenarios {\color{black}we describe} below are first generated using SUMO; then, the traces are input to ns-3 to simulate the location of the vehicular nodes. In SUMO, new vehicles are created at different times, which continuously enter and leave the DOCA during the simulated time. In ns-3, however, all nodes need to be created at the beginning of a simulation. Therefore, we turn off the radio functionality of any node outside the DOCA during the network simulations in ns-3.

In our evaluations, we consider two realistic environments denoted as ``E1'' and ``E2''. E1 has a single lane per direction, which obliges vehicles to drive in an ordered manner, thus representing a use case similar to platooning. Whereas, E2 has two lanes per direction, which yields more dynamic mobility due to the second lane allowing overtaking. We consider two DOCA lengths of $L\textsubscript{DOCA}=500$~m and $L\textsubscript{DOCA}=1000$~m for both environments. The vehicle arrivals to the DOCA follow a Poisson distribution with rate $0.4$/s (mean of $2.5$~s inter-arrival time) per direction {\color{black}as per the 3GPP evaluation assumptions~\cite{3gppTR36885}}. The vehicles follow a stochastic driving behavior by randomly varying their speeds based on the utilized car-following and lane-changing models \cite{sumoCarfollow}, \cite{sumoLane}, which depend on, e.g., average speed, road length, etc., hence making the mobility even more realistic in the evaluation environments. We vary the mean and variance of the vehicle speeds in both environments to create different loads of {\color{black}the vehicular} traffic over time and space. Specifically, we consider two scenarios in terms of vehicle density, denoted as loaded (``L'') and highly loaded (``HL''), both in E1 and E2, where the mean speed of {\color{black}the} vehicles is set to $120$~km/h and $50$~km/h, respectively (i.e., the slower, the denser). Further, the speeds among the vehicles are normally distributed, where we set the variance to $10\%$ and $30\%$ of the mean {\color{black}speed} values in E1 and E2, respectively. The higher variance of speeds in E2 increases the occurrence of vehicle take-overs across the two lanes.

Unless otherwise stated, the vehicles generate {\color{black}a} periodic V2V traffic with $T\textsubscript{msg}=100$~ms and $S\textsubscript{msg}=190$~B (as common to CAMs\cite{3gppTR36885}). We set the MCS index as $9$ and the number of RBs per subchannel as $16$ to fit the transmission of a single message of $190$~B into a single subchannel. In order to simulate loaded (and highly loaded) channel conditions in our evaluations, we assume that the resource pool consists of $2$ subchannels in the frequency domain (within an overall V2V bandwidth of $10$~MHz) and $10$~subframes in the time domain {\color{black}(hence denoted by $C^{2 \times 10}$)} unless otherwise stated, considering the number of vehicles and their V2V message generation rate. We accordingly set the length of the resource selection window of the mode 4 algorithm to $10$ ms with $T_1=4$~\cite{3gppTR36885} and $T_2=14$~ms. $ T_1 $ is to allow a processing time for the vehicles before they transmit their V2V messages, and $ T_2 $ sets a limit on the maximum latency of the transmissions. $P\textsubscript{keep}$ is set to $ 0 $, which leads to a dynamic re-selection of resources as much as possible, and has been shown to improve the reliability by avoiding persistent collisions especially under (highly) loaded channel conditions as in~\cite{wendland2019application}. To enable multiple collision domains within smaller DOCA lengths ($ 500 $~m), we set the V2V transmit power as $ -5 $~dBm, which yields a maximum communication range of around $ 200 $ m. This allows us to simulate environments with fewer vehicles, thus taking shorter simulation times. Nevertheless, we evaluate the performance of the algorithms also with the transmission power set to its allowed maximum value of $ 23 $~dBm \cite{3gppTS36101} in Section \ref{HighPower}. The further parameters related to the environment models, training of VRLS, and configuration of mode 4 are as listed in Table~\ref{Parameters}.

\section{VRLS Performance}
\label{Results}

\subsection{Reliability Performance}
\label{VRLS_reliability}
\begin{figure*}[h!]
	\begin{center}
		\subfigure[E1-L, $L\textsubscript{DOCA}=500$~m  \newline
		(speeds $\sim\mathcal{N}(120,12)$ km/h).]{\label{E1L500}\includegraphics[width=0.5\columnwidth]{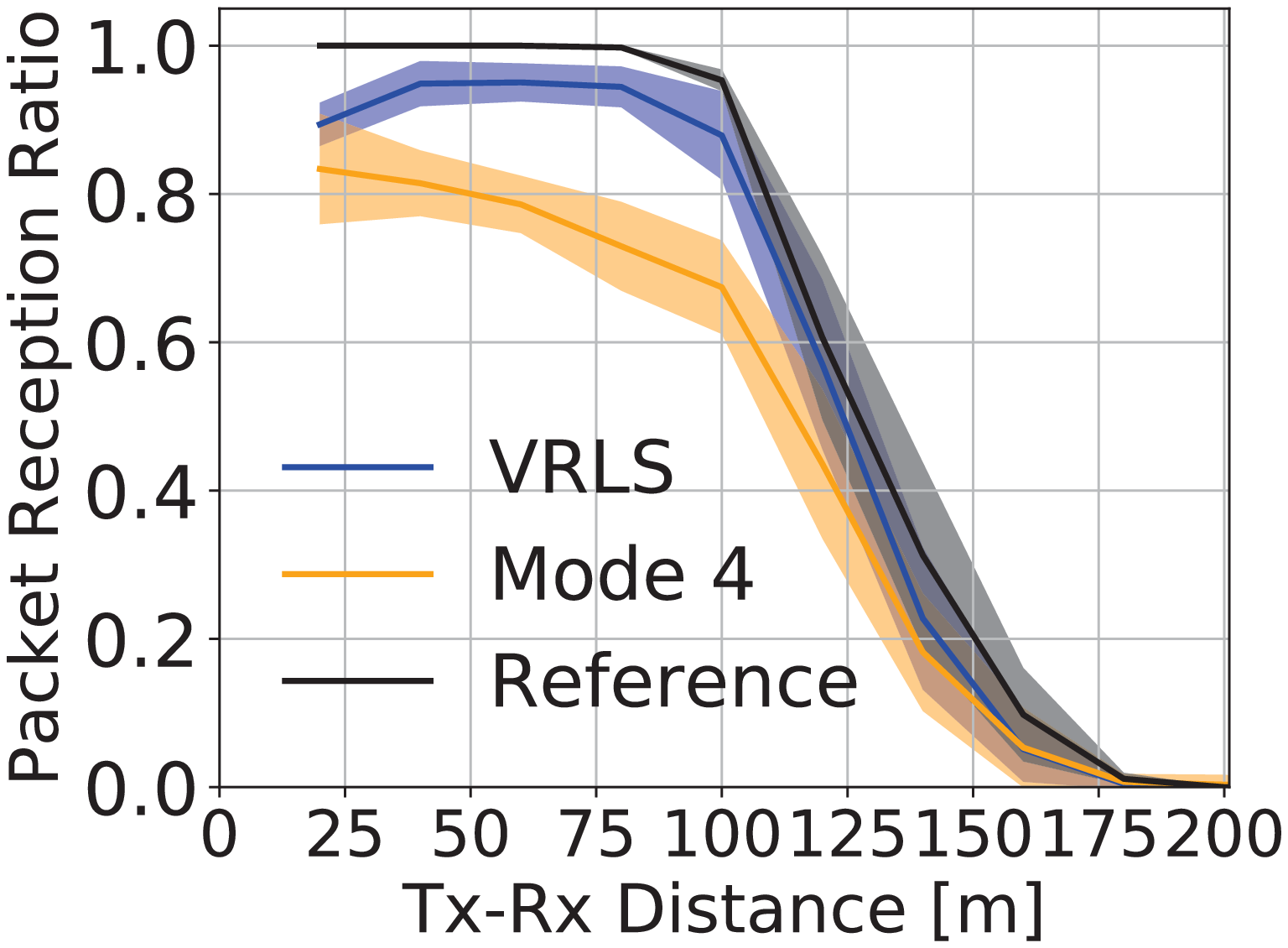}}  
		\subfigure[E1-L, $L\textsubscript{DOCA}=1000$~m \newline (speeds $\sim\mathcal{N}(120,12)$km/h).]{\label{E1L1000}\includegraphics[width=0.5\columnwidth]{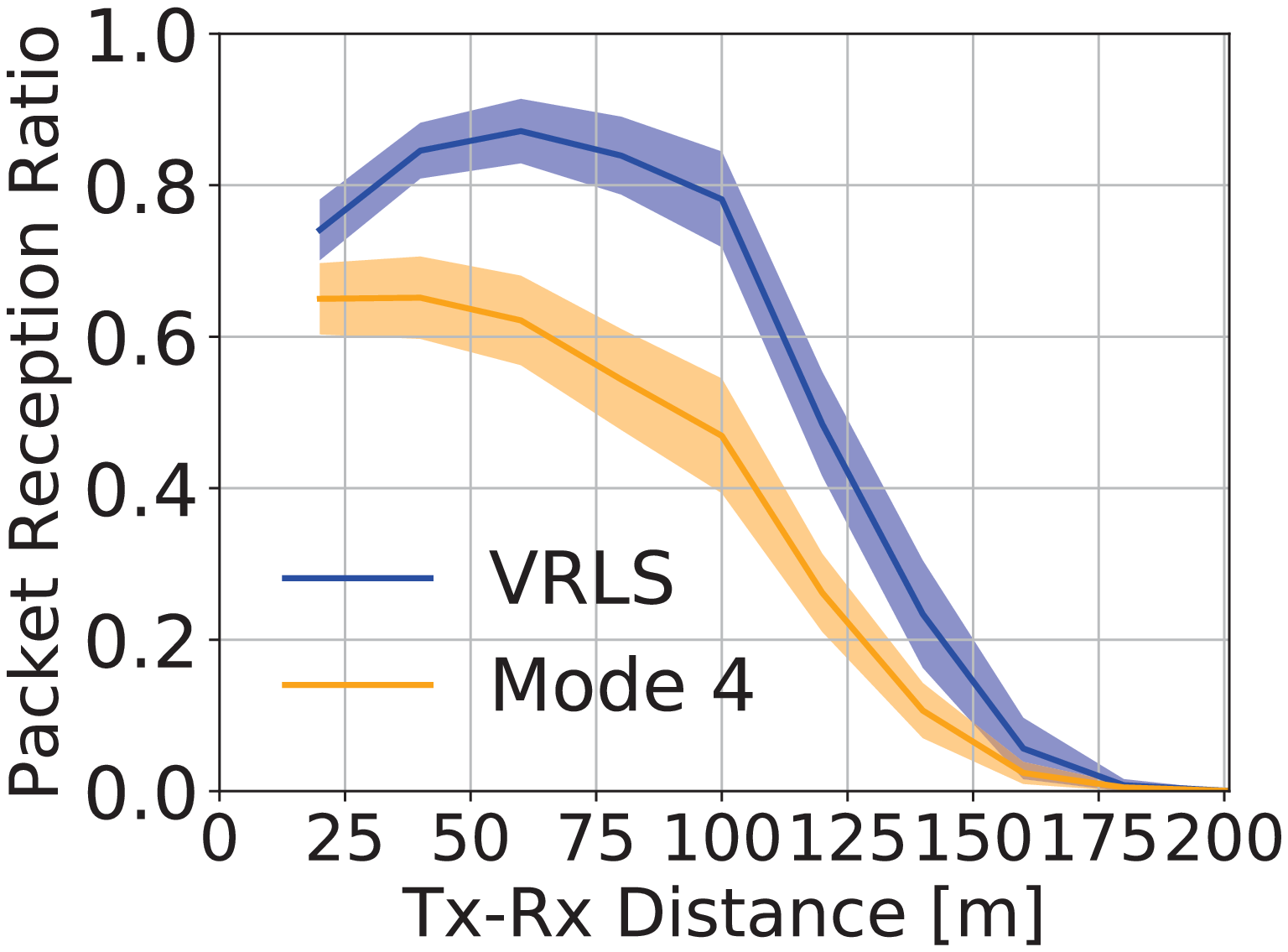}}
		\subfigure[E1-HL, $L\textsubscript{DOCA}=500$~m \newline (speeds $\sim\mathcal{N}(50,5)$ km/h).]{\label{E1HL500}\includegraphics[width=0.5\columnwidth]{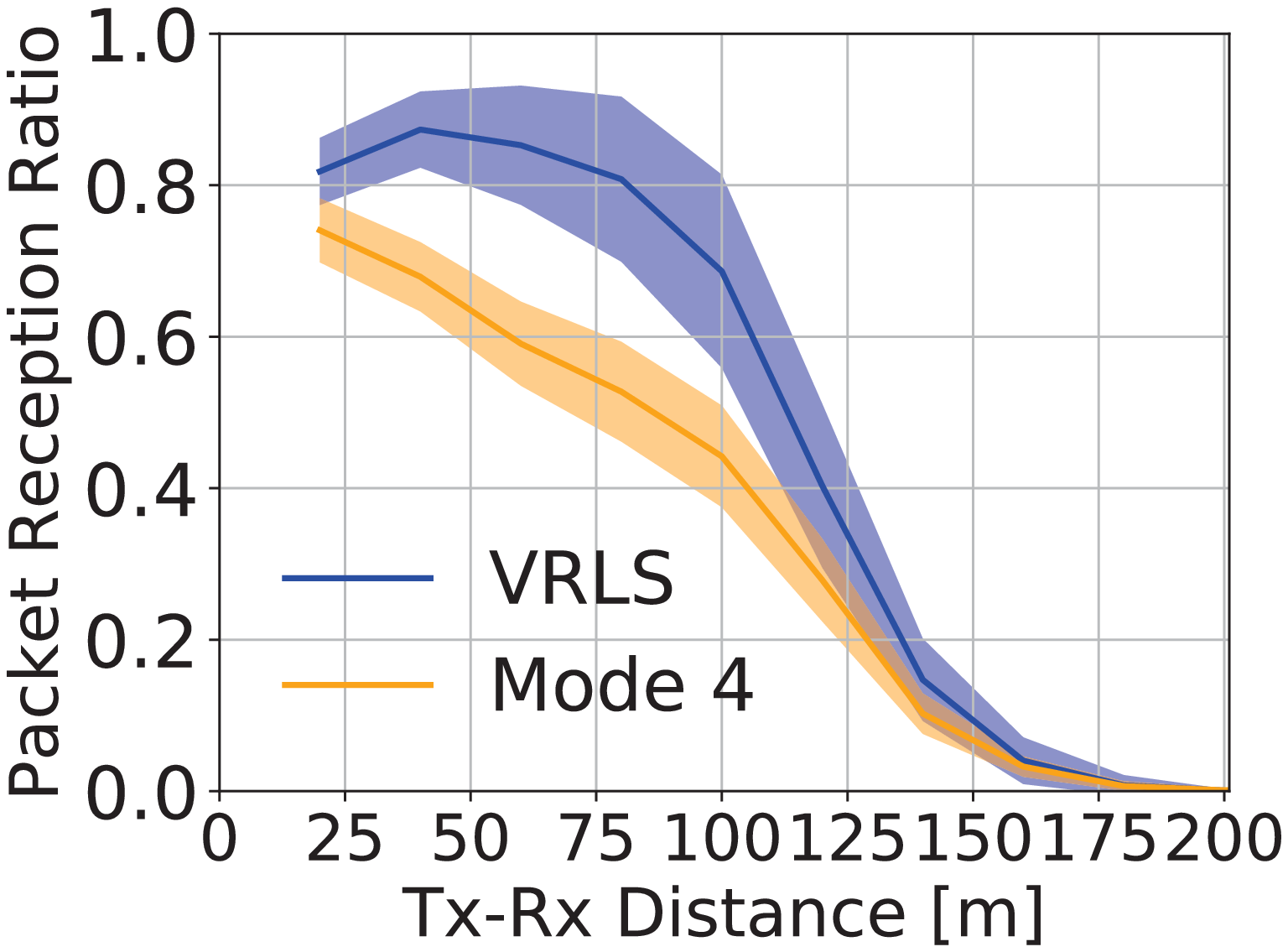}}
		\subfigure[E1-HL, $L\textsubscript{DOCA}=1000$m \newline (speeds $\sim\mathcal{N}(50,5)$ km/h).]{\label{E1HL1000}\includegraphics[width=0.5\columnwidth]{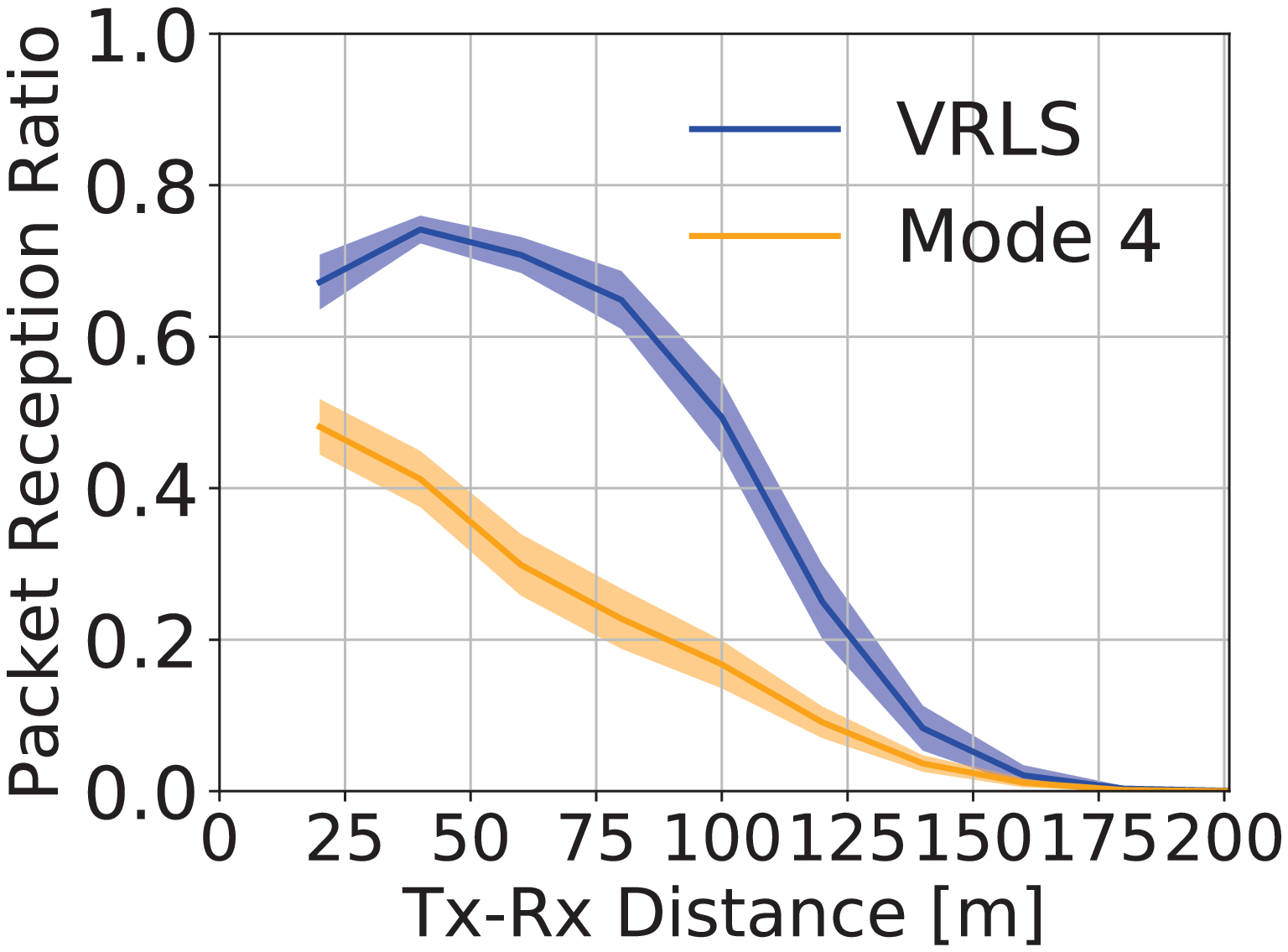}}
		\subfigure[E2-L, $L\textsubscript{DOCA}=500$~m \newline (speeds $\sim\mathcal{N}(120,36)$ km/h).]{\label{E2L500}\includegraphics[width=0.5\columnwidth]{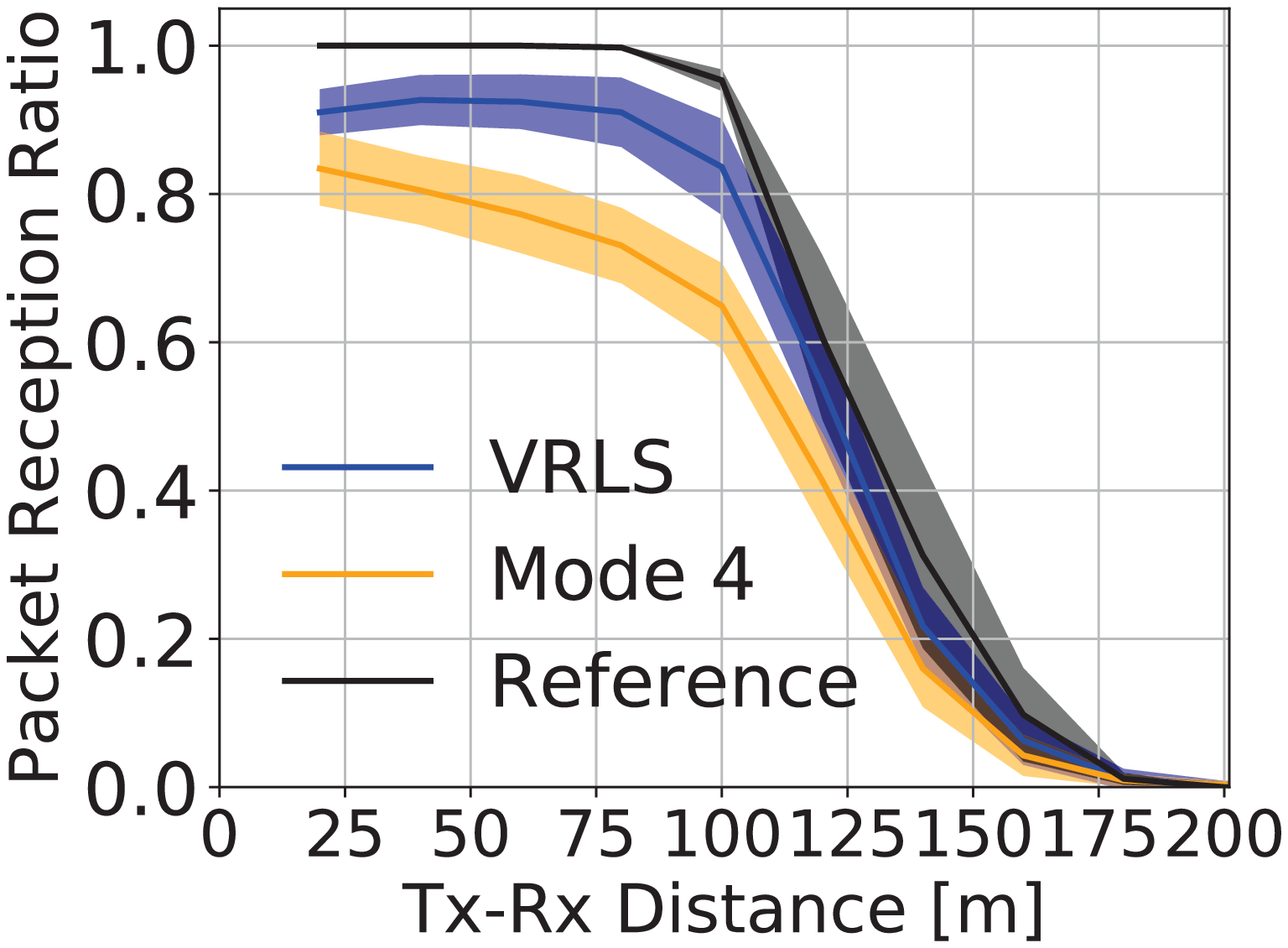}}
		\subfigure[E2-L, $L\textsubscript{DOCA}=1000$~m \newline (speeds $\sim\mathcal{N}(120,36)$km/h).]{\label{E2L1000}\includegraphics[width=0.5\columnwidth]{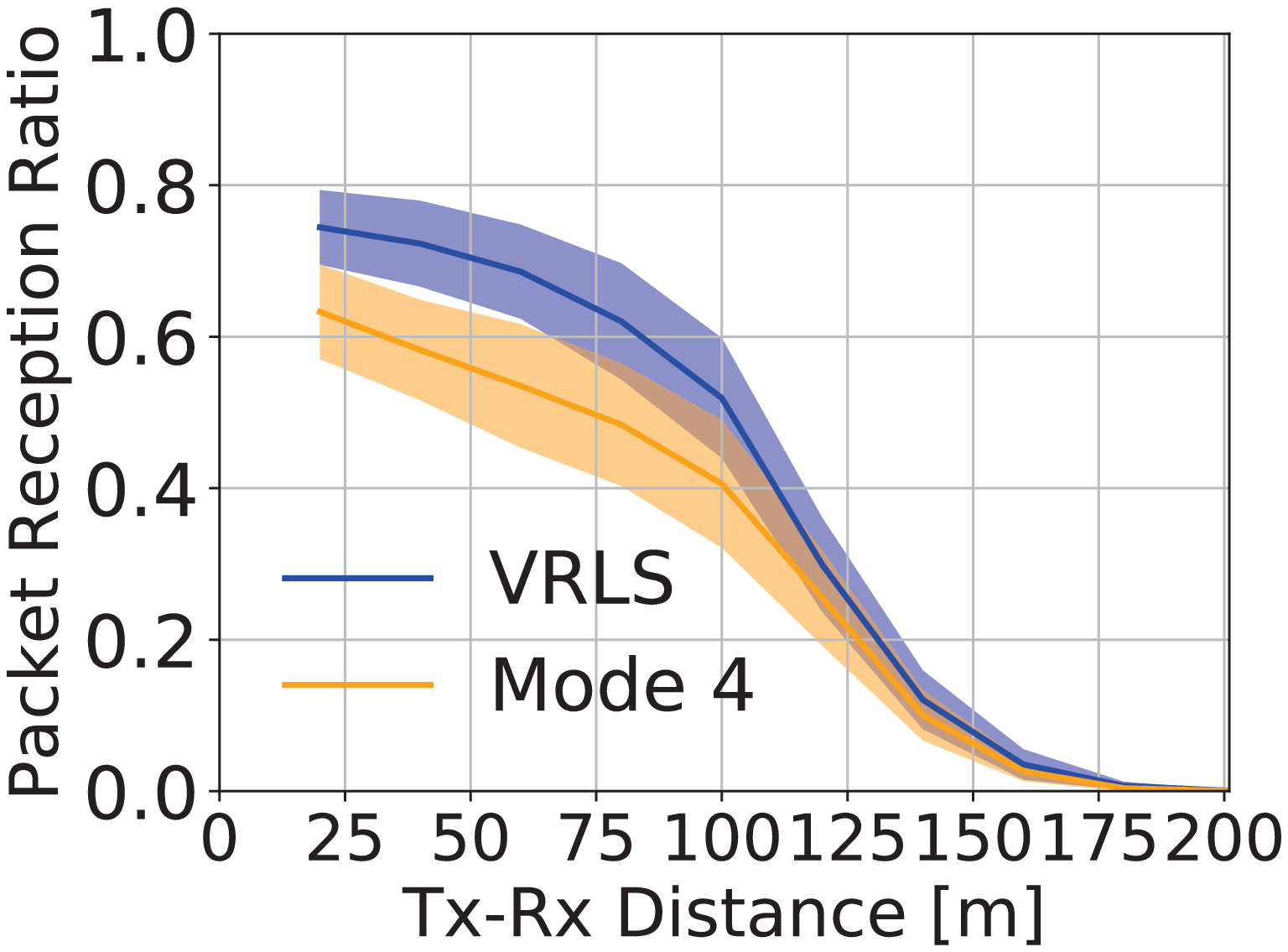}}
		\subfigure[E2-HL, $L\textsubscript{DOCA}=500$~m \newline (speeds $\sim\mathcal{N}(50,15)$ km/h).]{\label{E2HL500}\includegraphics[width=0.5\columnwidth]{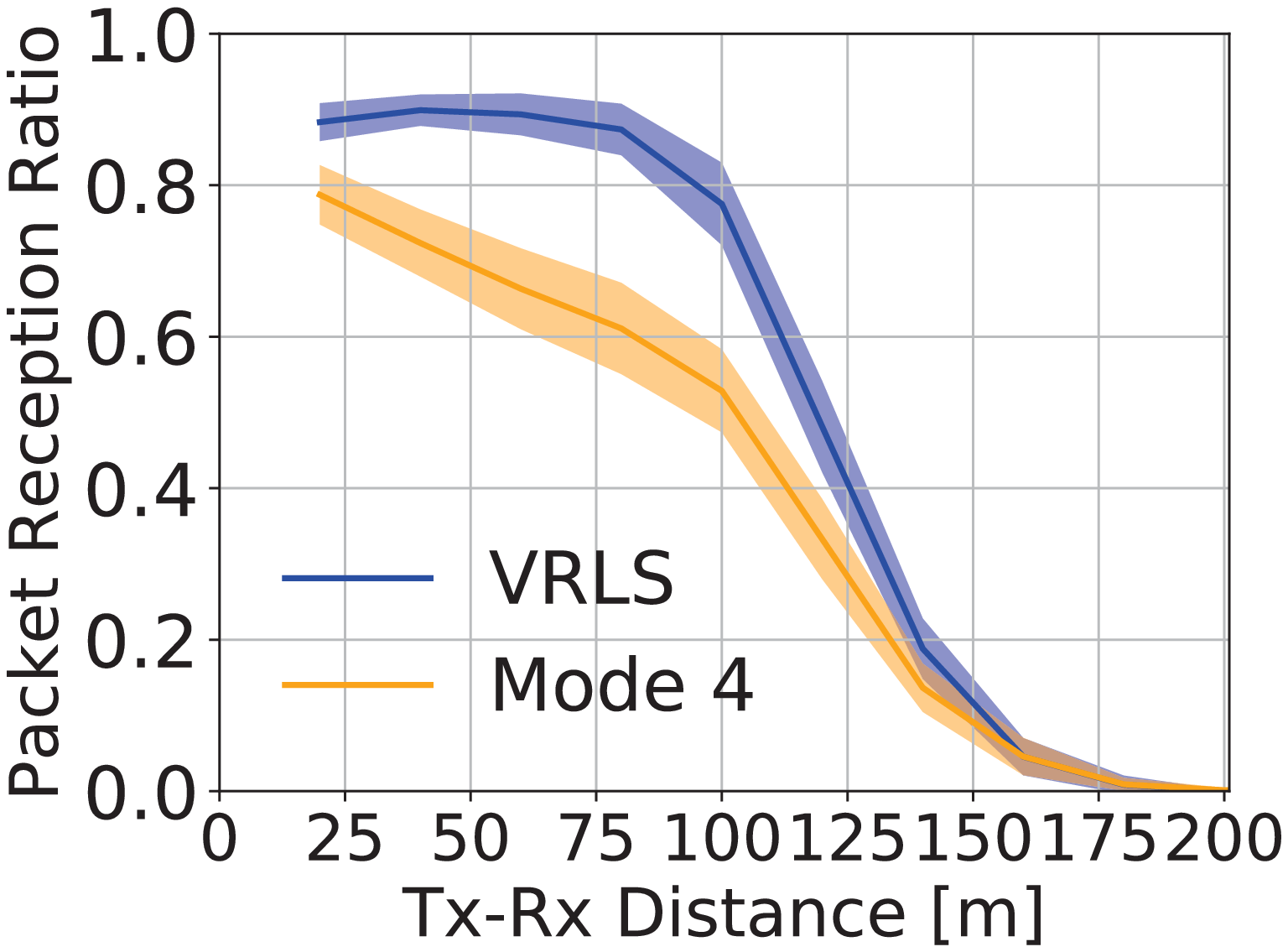}}
		\subfigure[E2-HL, $L\textsubscript{DOCA}=1000$m \newline (speeds $\sim\mathcal{N}(50,15)$km/h).]{\label{E2HL1000}\includegraphics[width=0.5\columnwidth]{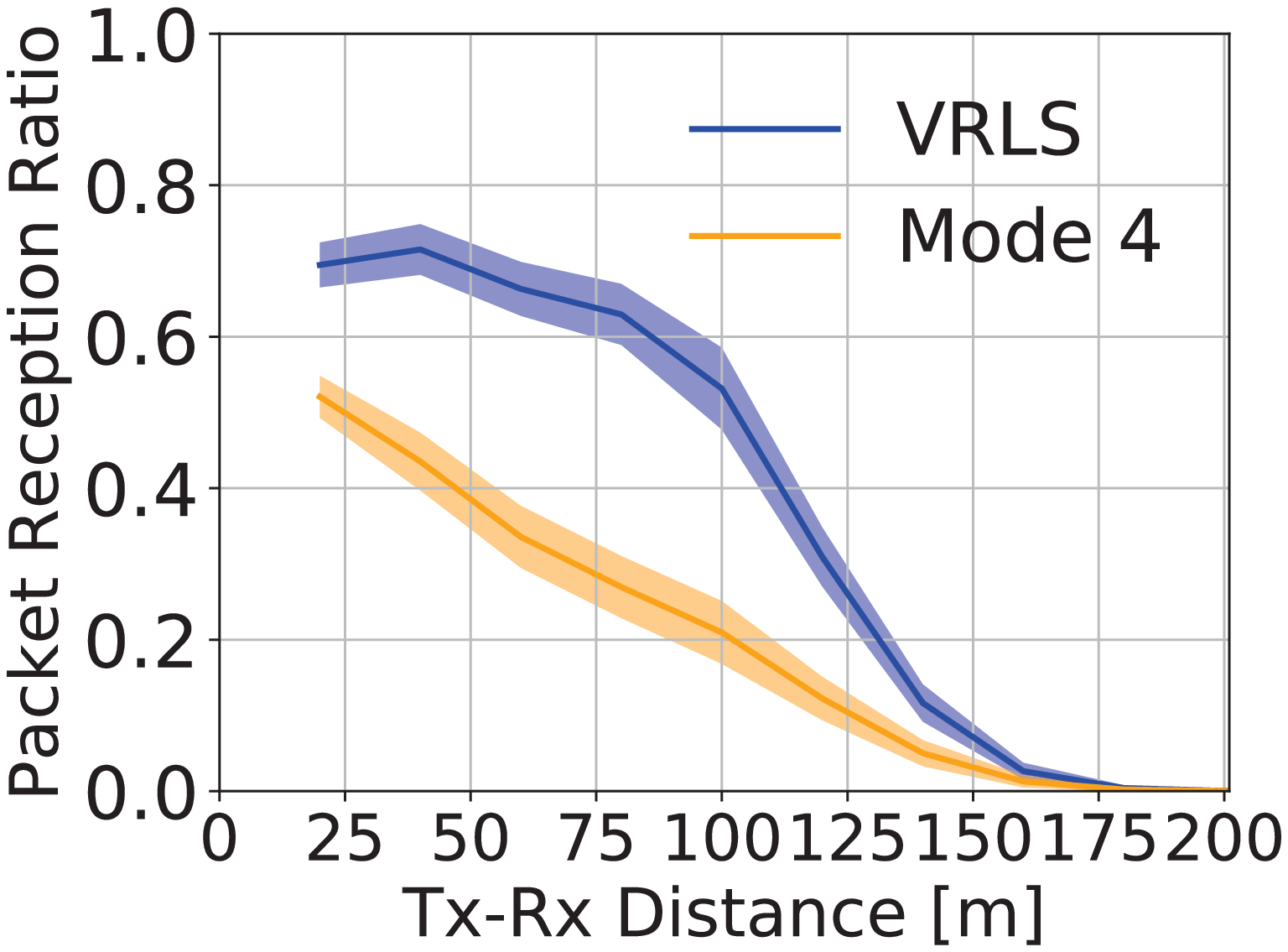}}
	\end{center}
	\vspace{-0.7\baselineskip}
	\caption{Performance of VRLS and distributed scheduling algorithm mode 4 on the DOCA environments E1 and E2 with different vehicular mobility scenarios. PRR vs Tx-Rx distance shown with mean (solid curve) and standard deviation (shade).}
	\vspace{-1\baselineskip}
	\label{E1}
\end{figure*}

In Fig. \ref{E1}, we compare the reliability of VRLS and mode~4 in E1 and E2 with loaded (L) and highly-loaded (HL) traffic {\color{black}with} two DOCA sizes of $L\textsubscript{DOCA}=500$~m and $1000$~m, using different subfigures. The plots provide the mean (solid curve) and the standard deviation (shaded region) of the average PRRs calculated in $10$ s intervals, for a simulation duration of $1000$~s (excluding the initial warm-up phase of $200$~s due to the initial random assignment of resources).

Fig. \ref{E1} shows that VRLS achieves better performance than mode 4 in all of the considered scenarios. VRLS is typically able to maintain a higher PRR over larger transmission ranges in both E1 and E2. The performance of mode 4 degrades more with the increasing distance between the transmitters and the receivers, mainly caused by the hidden-node problem leading to packet collisions. Beyond $100$~m, the path loss effect of the wireless channel becomes dominant, and inevitably reduces the PRR of both algorithms. 

To isolate the errors due to scheduling, in Table \ref{LossE1}, we numerically show the percentage of {\color{black}the} packet losses due to {\color{black}the} scheduling of the algorithms. The percentages at each transmitter-receiver (Tx-Rx) range are calculated as the difference between the achieved mean PRR and a reference value giving the maximum possible mean PRR in the environment. The reference values represent an ideal scenario, {\color{black}which assumes that} there are always sufficient resources for all transmissions, {\color{black}and the} packet losses {\color{black}are only due to} propagation errors. For convenience, we also plot the maximum possible PRR as a \textit{reference curve} in Fig. \ref{E1L500}, \ref{E2L500}, and in Fig. \ref{E2L1000_23dBm_Pool}. 

From Table \ref{LossE1}, we observe that VRLS has superior performance compared to mode 4 in all scenarios. Within {\color{black}a} $100$~m of Tx-Rx range, VRLS maintains a higher {\color{black}rate} of successful packets. Beyond this range, {\color{black}the} packet losses are predominantly caused by the propagation loss rather than {\color{black}the} scheduling, given the low transmit power. In scenarios E1-L and E2-L with $L\textsubscript{DOCA}=500$~m, VRLS shows {\color{black}a} performance close to the ideal scenario. The PRR for both algorithms is degraded considerably with the increased vehicular density, as well as the increased DOCA size that impact the interference conditions. In {\color{black}the} highly-loaded scenarios, {\color{black}the} collisions increase due to the allocation of the same resources to different vehicles. In such cases, although both algorithms perform sub-optimally given the limited {\color{black}number} of resources, VRLS results in half the packet losses compared to mode 4.

\begin{table*}[h!]
	\renewcommand{\arraystretch}{0.6} 
	\setlength{\tabcolsep}{0.5pt} 
	\caption{Percentages of Packet Loss due to Scheduling, and Mean Latency in E1 and E2}
	\vspace{-0.5\baselineskip}
	\label{LossE1}
		\centering
		\begin{tabular}{r|rr|rr|rr|rr|rr|rr|rr|rr}
			& \multicolumn{2}{c|}{\textbf{E1-L 500m}} & \multicolumn{2}{c|}{\textbf{E1-L 1000m}} & \multicolumn{2}{c|}{\textbf{E1-HL 500m}} & \multicolumn{2}{c|}{\textbf{E1-HL 1000m}} & \multicolumn{2}{c|}{\textbf{E2-L 500m}} & \multicolumn{2}{c|}{\textbf{E2-L 1000m}} & \multicolumn{2}{c|}{\textbf{E2-HL 500m}} & \multicolumn{2}{c}{\textbf{E2-HL 1000m}}\\
			
			Tx-Rx[m] & VRLS & Mode 4 & VRLS & Mode 4& VRLS & Mode 4 & VRLS & Mode 4 & VRLS & Mode 4 & VRLS & Mode 4& VRLS & Mode 4 & VRLS & Mode 4\\
			\hline
			\strut
			0-20    &10.6 & 16.6 & 25.9 & 35.0& 18.2 & 26.0 &32.8 & 51.9 &  9.0 & 16.6 &25.6 & 36.7& 11.7 & 21.3 &30.5 & 47.9\\
			20-40   & 5.1 & 18.6 & 15.4 & 34.8& 12.6 & 32.1 &25.8 & 58.8 &  7.3 & 19.5 &27.7 & 41.8& 10.1 & 27.6 &28.5 & 56.5\\
			40-60   & 5.0 & 21.4 & 12.8 & 37.8& 14.7 & 40.9 &29.2 & 70.1 &  7.6 & 22.7 &31.4 & 46.5& 10.6 & 33.7 &33.7 & 66.4\\
			60-80   & 5.3 & 26.8 & 15.8 & 45.4& 18.9 & 47.0 &34.9 & 77.0 &  8.7 & 26.7 &37.7 & 51.4& 12.4 & 38.7 &36.8 & 72.8\\
			80-100  & 7.4 & 27.9 & 17.2 & 48.4& 26.7 & 51.1 &46.0 & 78.6 & 11.7 & 30.4 &43.4 & 54.8& 17.8 & 42.5 &42.2 & 74.4\\
			100-120 & 3.6 & 17.1 & 12.2 & 34.5& 20.4 & 32.8 &35.6 & 51.6 &  6.5 & 19.4 &30.8 & 35.2& 12.6 & 27.4 &29.7 & 48.4\\
			120-140 & 8.5 & 13.1 &  7.9 & 20.6& 16.6 & 21.0 &22.9 & 27.6 &  9.5 & 15.3 &19.2 & 21.2& 12.5 & 17.6 &19.6 & 26.3\\
			\hline
			\strut
			Latency [ms] & 9.43 & 7.80 & 9.40 & 7.89 & 9.40 & 8.79 & 9.41 & 7.96 & 9.39 & 8.25 & 9.30 & 8.04 & 9.44 & 8.80 & 9.36 & 7.92\\
		\end{tabular}
		\vspace{-1\baselineskip}
		\label{E1table}
\end{table*}

We examined the policy that VRLS learned by observing the course of states and actions. VRLS develops a strategy to divide the resource pool dynamically into two directions of the highway, in proportion to the data traffic demand. Simultaneously, VRLS performs \textit{resource reuse per direction}, hence mitigating the hidden-node problem. Given the loaded conditions, HD errors in the network become inevitable even though the collisions could be avoided. Namely, vehicles can be allocated to different subchannels, yet sharing the same subframe. {\color{black}To illustrate, with $C^{2 \times 10}$, when there are $ 20 $ vehicles in the DOCA, each of the $ 10 $ subframes would be shared by two vehicles using different subchannels in order to avoid any resource collisions. Yet, such {\color{black}an} allocation would result in HD errors among these vehicles when they enter within each other's communication range. In fact, in such loaded scenarios, the probability of unsuccessful transmissions due to HD errors would be even larger if the pool consisted of fewer subframes and more subchannels (e.g., $C^{4 \times 5}$) as more vehicles would be required to use the different subchannels sharing each subframe to avoid any collision errors. We evaluate and observe such different resource pool configurations in Section~\ref{Results2}.} Yet, VRLS learned to assign the resources with HD conflicts, i.e., the subchannels sharing the same subframe, to vehicles in the \textit{opposite} directions rather than to the nearby vehicles in the same direction. Such assignment strategy results in comparatively fewer HD errors, as those vehicles pass by each other for a shorter duration of time. The outcome is especially observable in E1. The single-lane traffic in E1 results in a minimum inter-vehicle distance of about $40$~m within a lane. The Tx-Rx distances below $40$~m have a reduced PRR, which occurs only as a result of {\color{black}the} vehicles {\color{black}from opposite directions} passing by each other. Compared to E1, the second lane in E2 enables the vehicles to overtake each other, hence resulting in a more dynamic environment. Subsequently, the distance between the vehicles in the same direction can take any value, resulting in a smoother decrease of {\color{black}the} PRR with the increasing Tx-Rx range, as compared to E1. The trained VRLS policy is efficiently deployable in such a dynamic environment with varying vehicular density and network load over time and space, where it can deliver higher reliability than mode 4.

\subsection{Impact of Network Quality of Service on V2V Applications}
\label{NetworkToApp}

We evaluate the impact of PRR on the performance of V2V applications, by utilizing the awareness probability $P\textsubscript{A}$. As an illustrative example, the lane-change warning application requires at least $n=3$ messages to be received within $T=1$~s with $P\textsubscript{A}=99\%$ to make the neighboring vehicles aware of the intended maneuver \cite{An2011}. Following our assumption of $10$~Hz message frequency, i.e., $k=10$, and assuming independent message errors, this translates into a PRR requirement (i.e., $p$ given $P\textsubscript{A}$) of $61.12\%$. In our multi-lane environment E2, VRLS can achieve such {\color{black}a} PRR at up to {\color{black}a} $120$~m of range for $L\textsubscript{DOCA}=500$~m (see Fig. \ref{E2L500} and \ref{E2HL500}), and up to around $80$~m in $L\textsubscript{DOCA}=1000$~m (Fig. \ref{E2L1000} and \ref{E2HL1000}). In comparison, mode 4 achieves around $100$~m and $80$~m of {\color{black}an} awareness range in $L\textsubscript{DOCA}=500$~m under {\color{black}the} loaded and highly-loaded traffic, respectively. In the case of $L\textsubscript{DOCA}=1000$~m, mode 4 yields an awareness range of $30$~m for the loaded scenario, and cannot satisfy the requirement at all for the highly-loaded scenario.

\subsection{Performance under Aperiodic V2V Traffic}

\begin{figure}[!t]
	\centering
	\includegraphics[width=0.6\columnwidth]{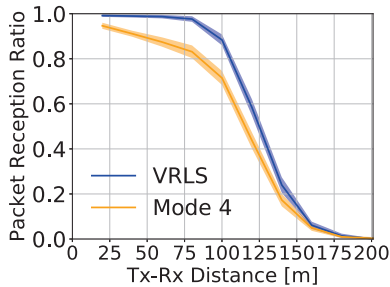}
	\vspace{-1\baselineskip}
	\caption{PRR evaluation of VRLS and mode 4 with aperiodic V2V traffic in the environment E2-HL with $L\textsubscript{DOCA}=1000$~m. PRR vs Tx-Rx distance shown with mean (solid curve) and standard deviation (shade).}
	\label{E2HL1000aper}
	\vspace{-1\baselineskip}
\end{figure}

We further evaluate the performance of VRLS under event-triggered V2V traffic. The vehicles are assumed to generate a message upon each event, where the event arrivals for each vehicle follow a Poisson distribution with a rate of {\color{black}$1$ event/s ($X\textsubscript{evt}=1/$s)}. In Fig. \ref{E2HL1000aper}, we report the PRR performance of the algorithms in scenario E2-HL with $L\textsubscript{DOCA}=1000$~m {\color{black}by considering aperiodic traffic only (not coexisting with periodic traffic)}. The event-triggered traffic results in less frequent V2V message generation as compared to the periodic traffic, which effectively creates a lower network load. Accordingly, the performance of both algorithms is increased (observed also in other scenarios), with VRLS achieving a PRR very close to $100\%$ up to a range of $80$~m. The results show that the policy learned by VRLS for {\color{black}the} periodic traffic is applicable to the aperiodic type of traffic as well. On the other hand, mode 4, which is a solution primarily designed for periodic V2V traffic, underperforms in this setting.

\subsection{{\color{black} Performance under High Transmit Power and Larger Pool}}
\label{HighPower}
For all of the results above, the vehicle transmit powers are set to $P\textsubscript{Tx}=-5$~dBm, whereas the allowed maximum for V2V transmissions is $23$~dBm \cite{3gppTS36101}. To ensure that the performance of VRLS holds for larger and arguably more realistic communication ranges, we evaluate the performance of the algorithms with the transmission power set to $23$~dBm for all vehicles in E2-L with $L\textsubscript{DOCA}=1000$~m. In order to compensate for the increased interference caused by {\color{black}the} high-powered transmissions, we consider a pool that consists of $2$ subchannels and $50$ subframes, i.e., $C^{2\times 50}$, which is five times larger than the resource pool configuration $C^{2\times 10}$ we have considered so far. For this scenario, VRLS is trained in E0 as well, but utilizing the resource pool $C^{2\times 50}$. 

\begin{figure}[!t]
	\centering
	\includegraphics[width=0.6\columnwidth]{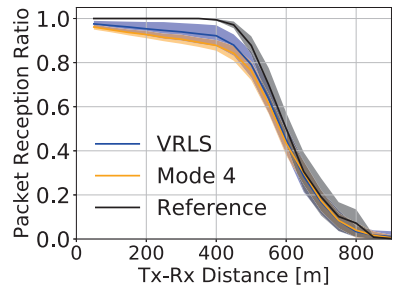}
	\vspace{-1\baselineskip}
	\caption{PRR vs Tx-Rx distance with mean (solid curve) and standard deviation (shade) for VRLS and mode 4 in comparison to the maximum possible value (reference) in E2-L with $L\textsubscript{DOCA}=1000$~m, $P\textsubscript{Tx}=23$~dBm, and $C^{2\times 50}$.}
	\label{E2L1000_23dBm_Pool}
	\vspace{-1\baselineskip}
\end{figure} 

Results of the algorithms are provided in Fig. \ref{E2L1000_23dBm_Pool}, where we see that both algorithms achieve very high reliability, close to {\color{black}a} $100$\% PRR at shorter Tx-Rx distances, owing to {\color{black}the} sufficiently provisioned resources. VRLS delivers {\color{black}a} marginally higher PRR than mode 4 at almost all Tx-Rx ranges. {\color{black} The results demonstrate that VRLS is trainable in environments having different resource pool configurations, and that the learned policy is applicable to scenarios with different transmission ranges.} The small percentage of {\color{black}the} packet losses results mainly from propagation errors, but also due to HD or even collisions to a small extent. In the case of mode 4, vehicles cannot sense the subframes they transmit on (due to the HD constraint); thus there exists a probability of selecting {\color{black}the} resources used by other vehicles that might interfere or collide. In the case of VRLS, although the learned policy avoids allocating the same resource to more than a single vehicle, it is challenging for the agent to learn the HD constraints in such a large and sparse state-space, where some of its assignments lead to HD errors. 

\begin{figure}[!t]
	\centering
	\includegraphics[width=0.6\columnwidth]{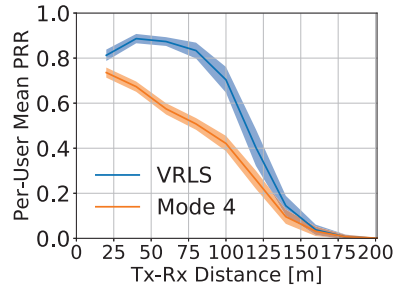}
	\vspace{-1\baselineskip}
	\caption{Per-user mean PRR of VRLS and mode 4 in E1-HL with $L\textsubscript{DOCA}=500$~m; shown with mean (solid curve) and standard deviation (shade).}
	\vspace{-0.7\baselineskip}
	\label{Fairness}
\end{figure}

\begin{figure}[!t] 
	\centering
	\includegraphics[width=0.9\columnwidth]{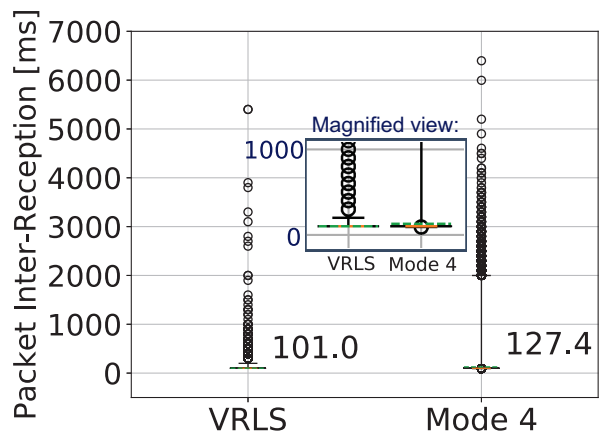}
	\vspace{-0.7\baselineskip}
	\caption{PIR performance of the algorithms in E2-HL with $L\textsubscript{DOCA}=500$~m, {\color{black}shown with mean (green, dashed, denoted): $ 101.0^V, 127.4^M$; median (orange): $ 100.0^V, 100.0^M$; $25$\textsuperscript{th} and $75$\textsuperscript{th} percentiles (black): both $ 100.0^V, 100.0^M$; $0.1$\textsuperscript{st} and $99.9$\textsuperscript{th} percentiles (whiskers): $ 100.0^V, 100.0^M$ and $ 200.0^V, 2000.0^M$; and outliers (rings), measured at Tx-Rx distances of $0-50$~m (V:VRLS, M:Mode 4, values in ms).}}
	\vspace{-0.5\baselineskip}
	\label{PIR}
\end{figure}

\subsection{User Fairness, Packet Inter Reception, {\color{black} and Latency}}
Although our solution gives an equal opportunity to all vehicles to transmit in the DOCA {\color{black}by allocating} resources, the PRR results do not provide the information on whether fairness is ensured, i.e., the PRR of a certain group of users is not sacrificed in favor of system-wide performance. In Fig.~\ref{Fairness}, we provide the mean and the standard deviation of the per-user average PRRs, to evaluate the variation of reliability across the users. The results are presented for the highly-loaded environment E1-HL with $L\textsubscript{DOCA}=500$~m. We observe that VRLS is able to deliver its performance without sacrificing user fairness. Both VRLS and mode 4 achieve a similar variation of mean PRR across the users, where the standard deviation is around $0.025$ considering all Tx-Rx ranges.

In Fig. \ref{PIR}, we report another \textit{per-user} performance metric, PIR, in the environment E2-HL with $L\textsubscript{DOCA}=500$~m, measured {\color{black}within a} $50$~m {\color{black}of} Tx-Rx range, in terms of mean and percentiles. We observe that VRLS achieves better performance than mode 4, the latter resulting in {\color{black}at least} $25\%$ larger PIR on average. Note that for both algorithms, all quartiles of the PIR are at $100$~ms, which is equal to $T\textsubscript{msg}$. We have observed that the relative PIR performance of the algorithms in the other scenarios are also similar, where VRLS achieves a mean PIR close to $100$~ms, at most reaching $106$~ms, and mode 4 resulting in mean values ranging from around $115$~ms up to $200$~ms {\color{black}that mainly increase} with the network load. For the high-transmit-power scenario in Fig. \ref{E2L1000_23dBm_Pool}, as the best case, VRLS and mode 4 can achieve {\color{black}a mean PIR} of $100.4$ and $100.6$ ms, respectively. Combined with our analysis in Section \ref{NetworkToApp}, it is evident that VRLS is able to provide superior awareness to the vehicular users while maintaining fairness, which benefits the V2V applications.

We report the mean latency measured in all evaluation scenarios in the last row of Table \ref{E1table}. Note that the messages are at least delayed by the processing time across the communication layers, which is assumed to be $4$ ms~\cite{3gppTR36885}, and at most delayed by the time-length of the resource pool, which is $10$ ms, plus the processing time, as all messages are scheduled within the utilized resource pool by both algorithms. Both VRLS and mode 4 yield a similar latency of around $9$ ms, on average.

\subsection{Handling Different Resource Pool Configurations}
\label{Results2}
	
The network might need to operate different resource pool configurations that may vary in terms of the number of resources in time and frequency, i.e., $K$ and $M$, so as to serve different V2V services with different traffic requirements, or depending on the availability of resources. Therefore, we are interested {\color{black}in studying} whether and how VRLS can handle a \textit{set} of different resource pools configured with different number of resources in time and frequency, with a \textit{single} training. Note that the pool configurations are usually a part of network planning; thus the configuration information would be available before operating the scheduler. Given this information, having a single policy that can achieve {\color{black}an} appropriate performance {\color{black}level} under all different configurations would be desirable, {\color{black}instead of} training multiple ones that can only operate on a specific configuration. That said, it is a challenge to learn and solve the HD and collision constraints of different pools at once by a single policy, as each has a different impact on the performance. Such an approach requires careful consideration of the state and the other RL components.
	
\begin{figure}[!t]
	\centering
	\includegraphics[width=\columnwidth]{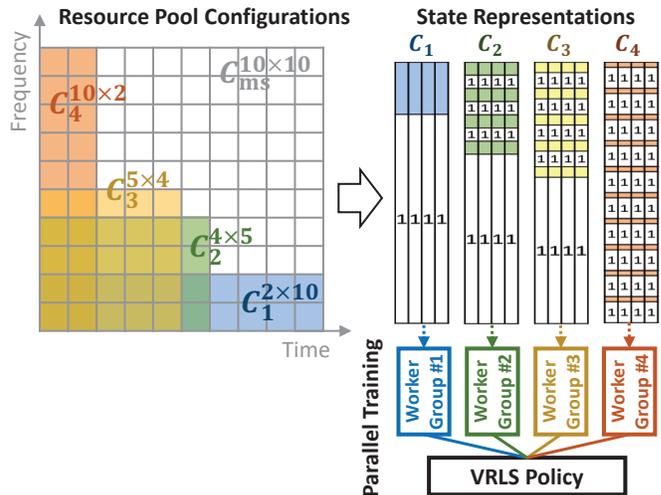}
	\vspace{-2\baselineskip}
	\caption{Training VRLS with multiple resource pool configurations in parallel.}
	\label{MultiplePools}
	\vspace{-1\baselineskip}
\end{figure}
	
\begin{figure}[!t]
	\centering
	\includegraphics[width=0.9\columnwidth]{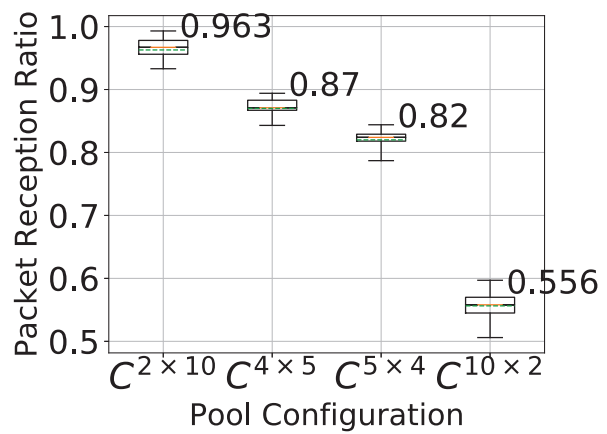}
	\vspace{-1\baselineskip}
	\caption{PRR performance of VRLS on different resource pool configurations $C^{K\times M}$ having $K$ subchannels and $M$ subframes, with mean (green, dashed, denoted), median (orange), $25$\textsuperscript{th} and $75$\textsuperscript{th} percentiles (box), and $5$\textsuperscript{th} and $95$\textsuperscript{th} percentiles (whiskers).}
	\vspace{-1\baselineskip}
	\label{Pools}
\end{figure}
	
{\color{black}We describe our method to train a single policy for multiple resource pool configurations as follows.} Consider a set of different resource pool configurations \{$C^{K_1\times M_1}_1$, $C^{K_2\times M_2}_2$, ...\} to be operated by the network, {\color{black}where} each pool $C_i$ consists of a different number of subchannels $K_i$ and subframes $M_i$. We first determine a superset (``master'') pool configuration $C^{K\textsubscript{ms}\times M\textsubscript{ms}}_{\text{ms}}$, which can accommodate any configuration in the set. Accordingly, the dimensions of $C\textsubscript{ms}$ are selected as $K_{\text{ms}}=\max(K_1,K_2,...)$ and $M_{\text{ms}}=\max(M_1,M_2,...)$. VRLS is provided with {\color{black}a} state- and {\color{black}an} action-space having the same number of resources as in $C\textsubscript{ms}$, i.e., $K\textsubscript{ms}\times M\textsubscript{ms}$. {\color{black}As illustrated in Fig. \ref{MultiplePools}, considering a case where the network operates four different resource configurations \{$C^{2\times 10}_1$, $C^{4\times 5}_2$, $C^{5\times 4}_3$, $C^{10\times 2}_4$\}, a master pool of $C^{10\times 10}_{\text{ms}}$ accommodates all four; thus the state- and the action-space of VRLS consist of $100$ resources. We provide {\color{black}the} different pool configurations to different groups of workers training the VRLS policy in parallel. Each group of workers $i$ is trained with the pool configuration $C_i$, {\color{black}where we only ``disclose'' the resources of $C_i$ within $S_t$ by replacing the rows corresponding to other resources with the row vector $[1, 1, 1, 1]$.} Further, if the worker selects such a resource, we provide a large negative reward and execute no actions until the worker selects a resource within its own pool. {\color{black}With such training, we aim at limiting the action selection of the policy only to the represented subset of the resources in $S_t$.}


\begin{figure}[!t]
	\centering
	\includegraphics[scale=0.26]{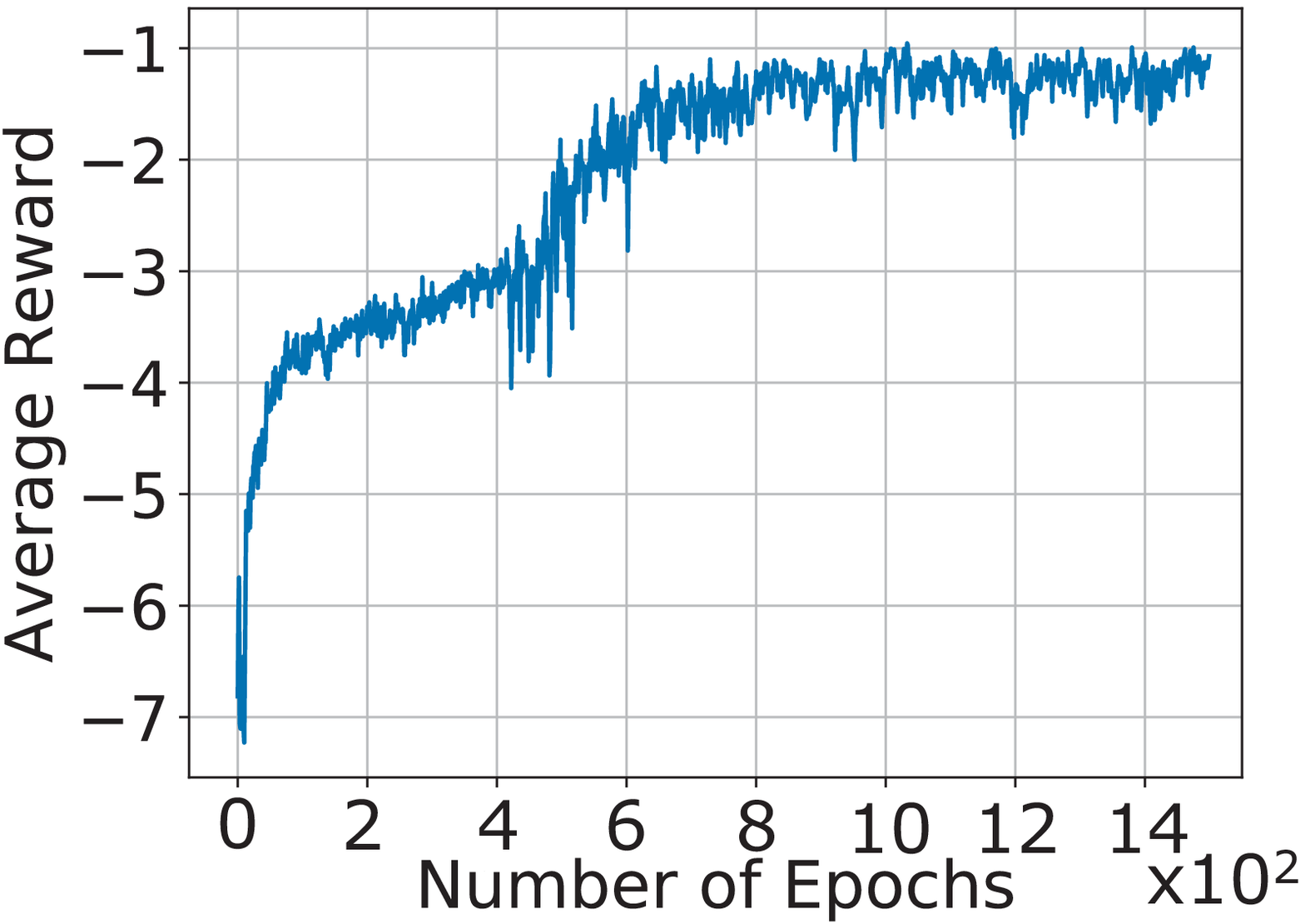}  
	\includegraphics[scale=0.27]{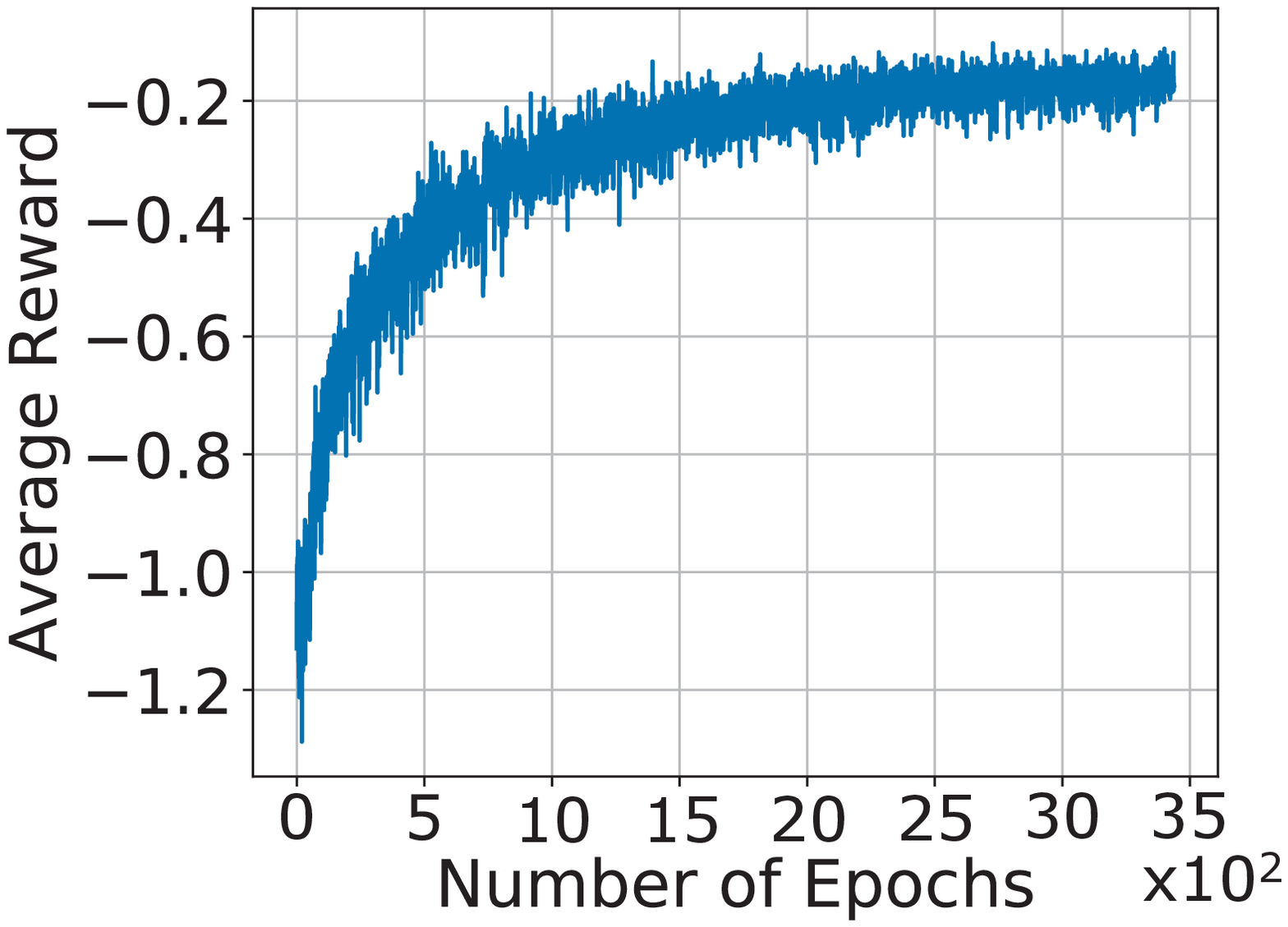}
	\includegraphics[width=0.49\columnwidth]{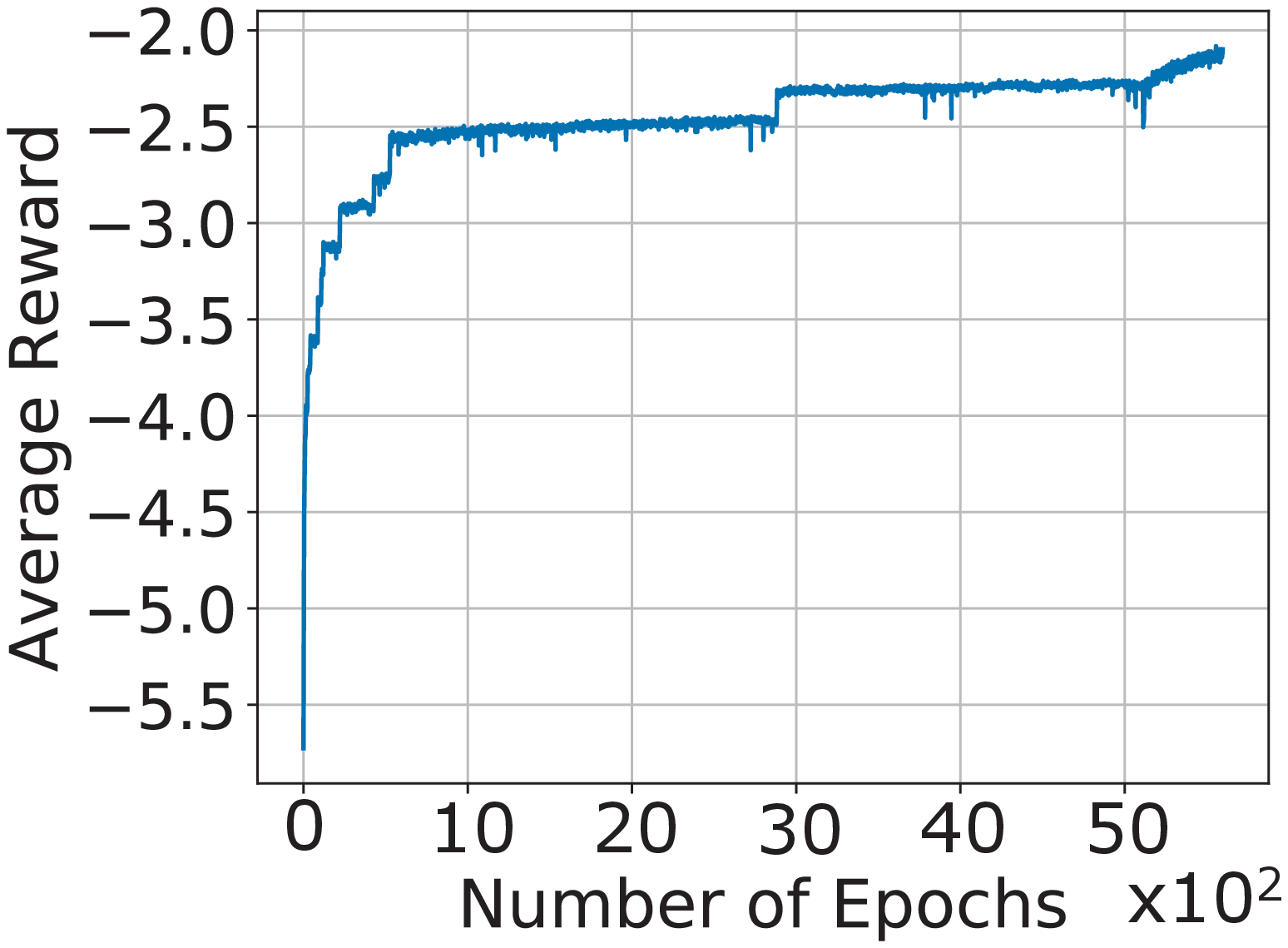}
		\vspace{-0.6\baselineskip}
	\caption{ {\color{black}Learning curves of VRLS in the training environment E0 with resource pool configuration $C^{2 \times 10}$ (top left), $C^{2 \times 50}$ (top right), and with multiple configurations \{$C^{2\times 10}_1$, $C^{4\times 5}_2$, $C^{5\times 4}_3$, $C^{10\times 2}_4$\} (bottom)}.}
	\vspace{-1.6\baselineskip}
	\label{Pretraining}
\end{figure}

In the following, we evaluate our solution {\color{black}for four different configurations \{$C^{2\times 10}_1$, $C^{4\times 5}_2$, $C^{5\times 4}_3$, $C^{10\times 2}_4$\}.} VRLS is trained from scratch with a total of $40$ workers in parallel, in four groups of 10 workers. Each group is provided with one of the four different configurations. If a worker selects a resource outside its configuration, a reward of $R_{t+1}=-10$ is provided. In turn, to compensate for the higher variance in the rewards, {\color{black}the} training epoch length is increased to $200$ actions. We evaluate VRLS in a DOCA similar to E0, with $10$ vehicles initially placed on the highway with transmission range $R\textsubscript{Tx}=500$~m, resulting in a single collision domain (transmissions using the same TB are assumed to collide). Such a simple setting enables a deterministic calculation of performance bounds and better evaluation of whether the learned policy can deal with the constraints of {\color{black}the} different resource pools. Namely, in the case of $C^{2\times 10}_1$, when all $10$ vehicles reside in the DOCA, {\color{black}a} $100\%$ PRR would be achievable only if vehicles were assigned to TBs in different subframes. With $C^{4\times 5}_2$, the allocation for all $10$ vehicles would be optimal if all TBs were assigned orthogonally first in time, then in frequency. Every transmission would be received by all other vehicles except the one transmitting in the same subframe, due to the HD constraint. Thus, the best assignment of TBs would result in eight successful receptions out of nine receiving vehicles, yielding {\color{black}an} $88.\bar8\%$ PRR. Similarly, in the case of $C^{5\times 4}_3$ and $C^{10\times 2}_4$, when all of $10$ vehicles exist in the DOCA at the same time, orthogonal assignment of TBs first in time and then in frequency would yield $82.\bar2\%$ and $55.\bar5\%$ PRR, respectively.

In Fig. \ref{Pools}, we report the performance of VRLS when applied to the network with different resource pool configurations, in terms of {\color{black}the} PRR measured up to {\color{black}a} $500$~m of Tx-Rx range. We observe that VRLS yields {\color{black}a} mean PRR almost equal to the calculated bounds of $1.0$, $0.\bar8$, $0.8\bar2$, and $0.\bar5$ for the configurations $C^{2\times 10}_1$, $C^{4\times 5}_2$, $C^{5\times 4}_3$, and $C^{10\times 2}_4$, respectively. Note that {\color{black}the} larger PRRs are reached when fewer than the maximum of $10$ vehicles reside in the DOCA. VRLS is able to achieve such performance by learning a single policy that can handle distinct constraints of HD and collisions for different pools simultaneously.

\section{Learning Performance}
\label{Curves}
{\color{black}We provide the learning curves of VRLS in the training environment E0 with different resource pool configurations that we considered throughout our evaluations, in Fig. \ref{Pretraining}.} The curves represent the average reward collected by the trained workers versus the number of training epochs. VRLS converged to a stable performance level after around $1000$ epochs when trained with a single resource configuration $C^{2 \times 10}$ {\color{black}as per Section~\ref{VRLS_reliability}}. With the larger resource pool of $C^{2 \times 50}$ {\color{black}as per Section~\ref{HighPower}}, it took around two times longer for the agent to converge. This is because of additional exploration required by the increased state and the action space. VRLS obtained a larger average reward in the case of $C^{2 \times 50}$ owing to the sufficiently provisioned resources in the network. When VRLS is trained with four resource pool configurations in parallel, i.e., with \{$C^{2\times 10}_1$, $C^{4\times 5}_2$, $C^{5\times 4}_3$, $C^{10\times 2}_4$\} as per Section~\ref{Results2}, it took a longer time for the algorithm to converge as compared to the training with a single pool configuration. This is due to the different collision and HD constraints posed by each different resource pool configuration that VRLS needs to learn, as well as the larger state space, which results in slower convergence. The collected reward is smaller as it represents the average of dissimilar performance levels on {\color{black}the} different pools reported in Fig. \ref{Pools}. The overall performance is largely converged, which could be yet further optimized, such as via exhaustive training on the desired configuration, or with a larger number of workers, however, calling for increased training time and resources.

\vspace*{-0.3\baselineskip}

\section{Conclusions and Outlook}
\label{Conclusion}
We designed VRLS, a centralized scheduling approach for V2V communications outside the network coverage, based on RL. VRLS is applicable to a variety of vehicular environments having different sizes, densities, mobility, network load, wireless conditions, and resource configurations. Because VRLS can be pre-trained using simulations, it can be trained on a wider range of environments and resource configurations than what would be practically doable in the real world. In terms of reliability, VRLS outperforms the state-of-the-art mode 4 scheduler by reducing the packet loss by half in case of overloaded network conditions, and performing very close to the maximum possible level under low load. Furthermore, while achieving similar fairness and latency as mode 4, VRLS provides higher awareness among the vehicles.

Nevertheless, the actions of VRLS, namely assigning a single resource to each vehicle going outside the coverage, might become unfeasible under certain conditions. There could occur some situations such as road congestion due to an accident, or cases for which the agent might not be trained. Our complementary work \cite{sahin2019hybrid} addresses such issues by encompassing a \textit{hybrid} solution that combines the centralized RL-based approach with the distributed sensing-based scheduling for areas outside the network coverage. In any case, if the performance degradation is not tolerable, i.e., the communications cannot satisfy the requirements of a given V2V use case, the network could over-provision the resources, as we have shown, or congestion control mechanisms could be applied, considering such unforeseeable conditions. 

In another line of our work \cite{sahin2021ivrls}, we proposed \textit{iVRLS} (in-coverage VRLS) for scheduling \textit{in-coverage} V2V communications, which extends the VRLS design by taking advantage of resource assignments that are possible at all times and making use of instantaneous and exact knowledge of vehicular mobility. iVRLS is shown to improve the V2V transmission reliability under high traffic load, with less frequent scheduling as compared to a state-of-the-art scheduling algorithm. Further work remains to evaluate the network efficiency of the proposed algorithms, by measuring the number of resources they demand to satisfy target reliability requirements under varying conditions of network coverage.

%
%


\ifCLASSOPTIONcaptionsoff
  \newpage
\fi

\bibliographystyle{IEEEtran}
\bibliography{references}

%
%


\vspace*{-2\baselineskip}
\begin{IEEEbiography}[{\includegraphics[width=1in,height=1.25in,clip,keepaspectratio]{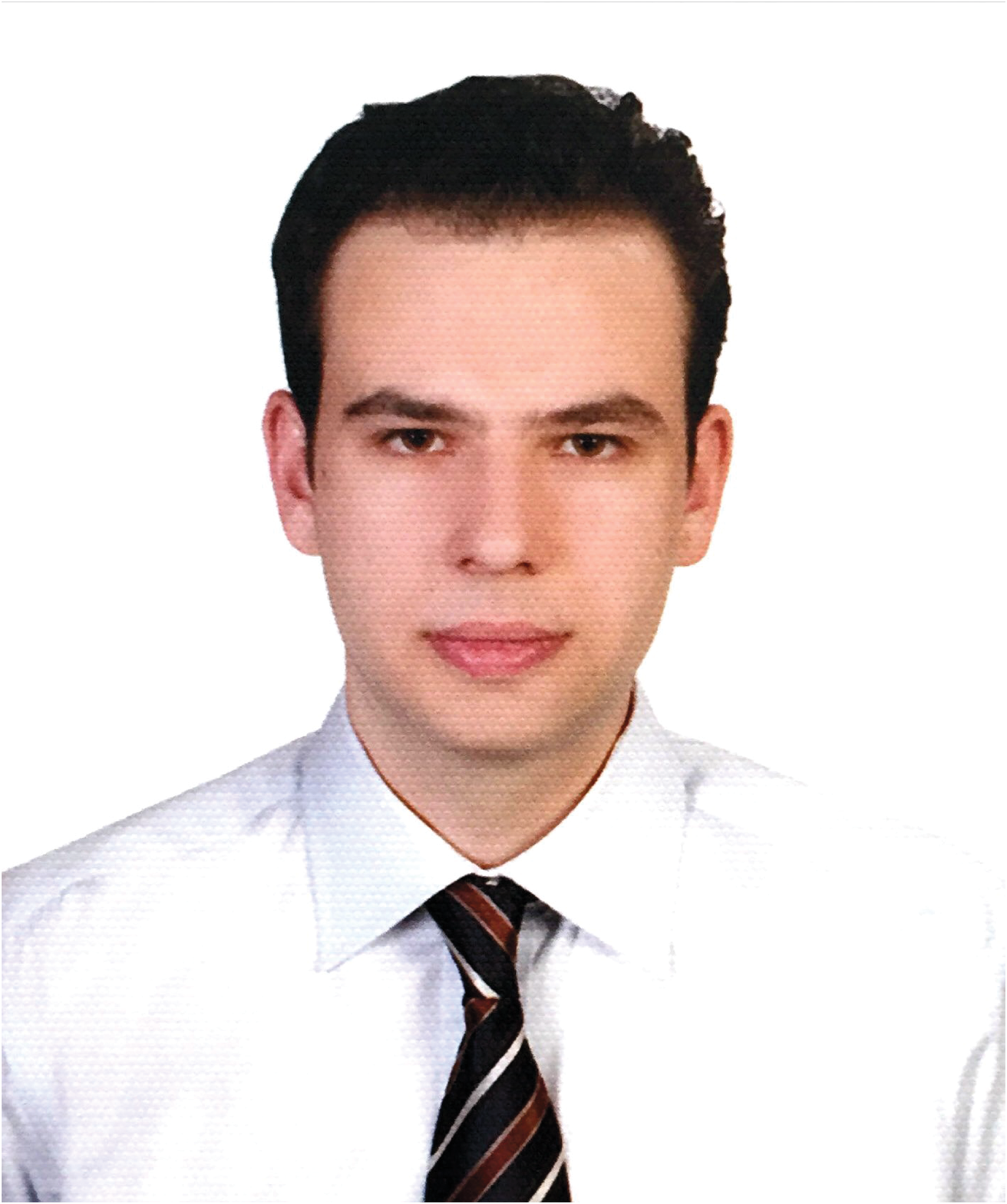}}]{Taylan \c{S}ahin} received the B.S. and M.Sc. degrees from the Middle East Technical University, Turkey, in 2014, and from the Technical University of Munich, Germany, in 2016, respectively. He is currently pursuing the Doctoral degree with the Technical University of Berlin, Germany. He was with Huawei Munich Research Center, Germany as a Research Student from 2015 to 2020. Since 2020, he has been with Nokia Standards, Munich. His master's work received the Best Paper Award, and he has served as a Co-Chair and TPC Member in various IEEE conferences and workshops. His research interests include resource allocation, positioning, and machine learning in wireless mobile networks.
\end{IEEEbiography}

\vspace*{-2\baselineskip}

\begin{IEEEbiography}[{\includegraphics[width=1in,height=1.25in,clip,keepaspectratio]{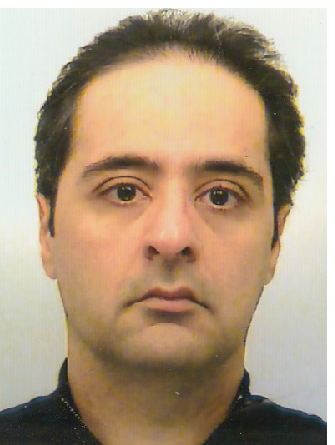}}]{Ramin Khalili} received his B.Sc. from Shiraz University, his M.Sc. from the Sharif University of Technology, both in Iran, and his Ph.D. from UPMC, France. He was with the UMASS Amherst, EPFL, and the TLabs in Berlin, before joining the Huawei Research Center in Munich, Germany. Ramin has published over fifty scientific papers on topics related to wireless networking, optimization, and machine learning and has received multiple best paper awards during these years.
\end{IEEEbiography}

\vspace*{-2\baselineskip}

\begin{IEEEbiography}[{\includegraphics[width=1in,height=1.25in,clip,keepaspectratio]{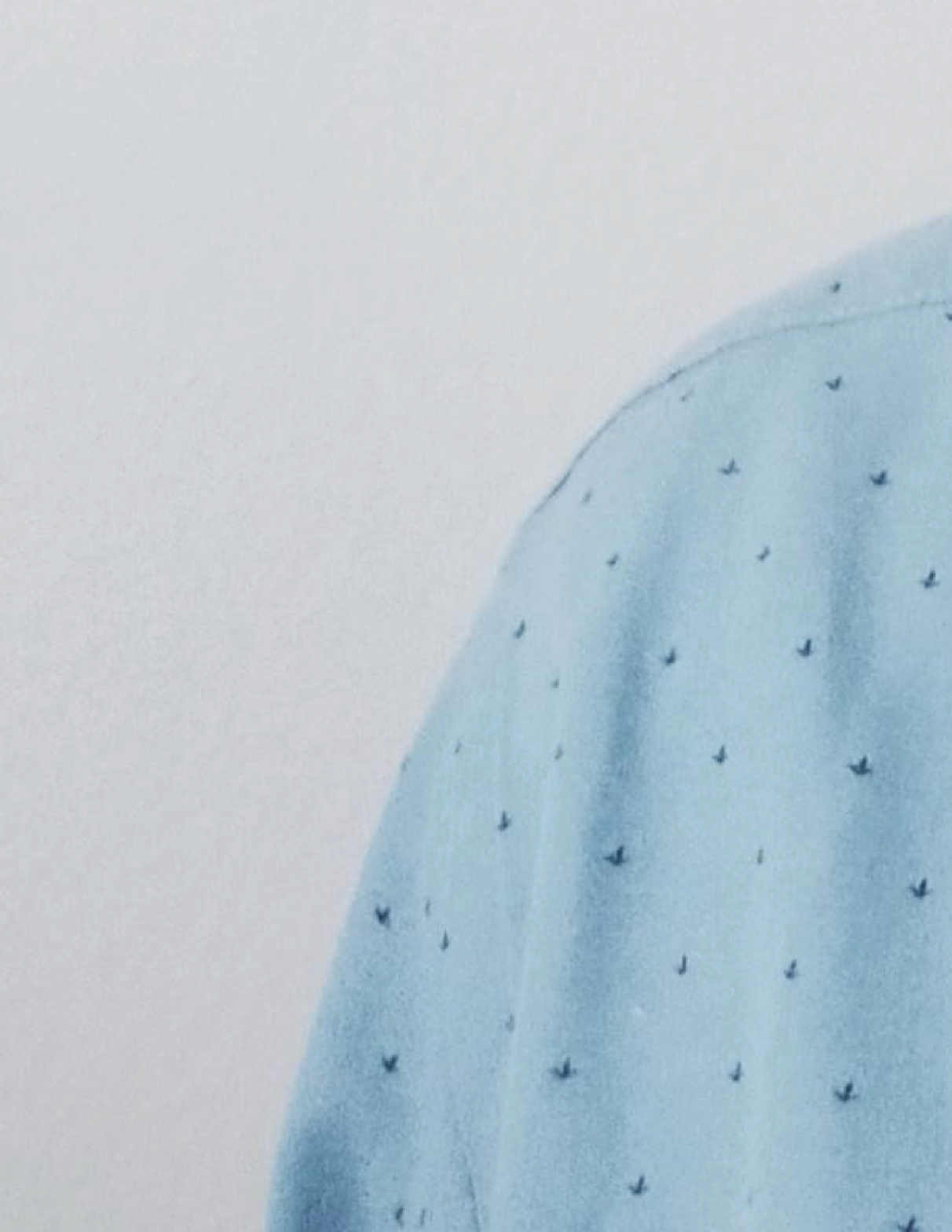}}]{Mate Boban [SM’21]} received his Ph.D. degree in electrical and computer engineering from Carnegie Mellon University. He is with Huawei Technologies, Munich Research Center, Germany. He is currently serving as an Associate Editor for IEEE Transactions on Mobile Computing. He has co-authored three papers that have received Best Paper Awards. His current research interests include channel modeling, resource allocation, and machine learning applied to wireless communications.
\end{IEEEbiography}

\vspace*{-2\baselineskip}

\begin{IEEEbiography}[{\includegraphics[width=1in,height=1.25in,clip,keepaspectratio]{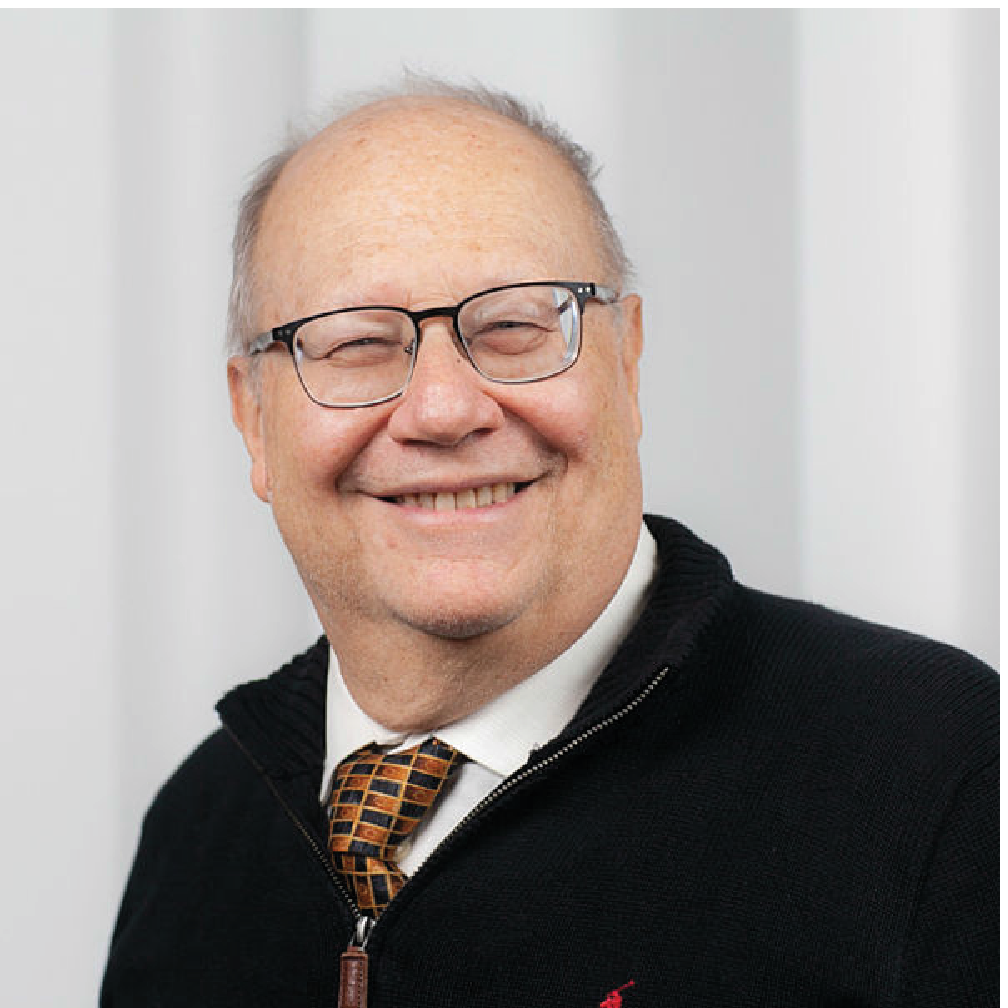}}]{Adam Wolisz [SM’99]} received the M.S., Ph.D., and Habilitation degrees from the Silesian University of Technology, Gliwice, Poland, in 1972, 1976, and 1983, respectively. After research work with the Institute of Complex Control Systems, Polish Academy of Sciences and GMD Fokus (Berlin) he was appointed Professor at the Technische Universitaet Berlin leading the Telecommunication Networks Group (1993-2018) and acted as Executive Director of the Institute for Telecommunication Systems (2001-2018). In parallel, he acted as adjunct professor at EECS, UC Berkeley (2005-2017) associated with Berkeley Wireless Research Center. Since retirement (2018) he has been active as Fellow at the Einstein Center Digital Future, Berlin.
\end{IEEEbiography}


\end{document}